\documentclass[a4paper,onecolumn,11pt]{quantumarticle}
\pdfoutput=1
\usepackage[utf8]{inputenc}
\usepackage[english]{babel}
\usepackage[T1]{fontenc}
\usepackage{amsmath}
\usepackage{amsfonts}
\usepackage{hyperref}
\usepackage{algorithm}
\usepackage{algpseudocode}
\usepackage{tikz}
\usepackage{lipsum}
\usepackage{graphicx}
\usepackage{tikz}
\usetikzlibrary{quantikz}
\usepackage{subcaption}
\usepackage{physics}
\usepackage[normalem]{ulem}
\usepackage{color,soul}

\usepackage[normalem]{ulem}

\DeclareMathOperator*{\argmax}{argmax}
\algrenewcommand\algorithmicrequire{\textbf{Input:}}
\algrenewcommand\algorithmicensure{\textbf{Output:}}

\begin{document}

\title{Automated Quantum Circuit Design with Nested Monte Carlo Tree Search}

\author[1]{Pei-Yong Wang}
\orcid{0000-0002-0665-6639}
\email{peiyongw@student.unimelb.edu.au}


\author[2,3]{Muhammad Usman}
\orcid{0000-0003-3476-2348}

\author[1]{Udaya Parampalli}
\orcid{0000-0002-9798-0134}

\author[2]{Lloyd C. L. Hollenberg}
\orcid{0000-0001-7672-6965}

\author[1,4]{Casey R. Myers}
\orcid{0000-0002-8838-7523}

\affil[1]{School of Computing and Information Systems, Faculty of Engineering and Information Technology, The University of Melbourne, Melbourne VIC 3010, Australia}
\affil[2]{School of Physics, The University of Melbourne, Parkville, VIC 3010, Australia}
\affil[3]{Data61, CSIRO, Clayton, Victoria, Australia}
\affil[4]{Silicon Quantum Computing Pty Ltd., Level 2, Newton Building, UNSW Sydney, Kensington, NSW 2052, Australia}

\maketitle

\begin{abstract}


Quantum algorithms based on variational approaches are one of the most promising methods to construct quantum solutions and have found a myriad of applications in the last few years. Despite the adaptability and simplicity, their scalability and the selection of suitable ans\"atzs remain key challenges. In this work, we report an algorithmic framework based on nested Monte-Carlo Tree Search (MCTS) coupled with the combinatorial multi-armed bandit (CMAB) model  for the automated design of quantum circuits. Through numerical experiments, we demonstrated our algorithm applied to various kinds of problems, including the ground energy problem in quantum chemistry, quantum optimisation on a graph, solving  systems of linear equations, and finding encoding circuit for quantum error detection codes. Compared to the existing approaches, the results indicate that our circuit design algorithm can explore larger search spaces and optimise quantum circuits for larger systems, showing both versatility and scalability.

\end{abstract}
\section{Introduction}\label{intro}

The variational quantum circuit (VQC, also known as parameterised quantum circuit, PQC) approach, first proposed for solving the ground state energy of molecules \cite{peruzzo2014variational}, have been extended to many open research problems including in the field of quantum machine learning \cite{schuldpetruccione2021}, quantum chemistry \cite{RevModPhys.92.015003}, option pricing \cite{2020optionpricing} and quantum error correction \cite{johnson2017qvector, Xu2021-dt}. The performance of VQC methods largely depend on the choice of a suitable ans\"atze, which is not an easy task because generally the search space is very large and it is not well established whether there is a common principle for designing such ans\"atze. For problems involving physical systems such as in quantum chemistry, we can rely on the well-defined properties of molecular systems for ans\"atz designing, like the  hardware efficient ans\"atze~\cite{2017hardwareefficientvqe} and physical-inspired ans\"atze, such as $k$-UpCCGSD~ \cite{physicalinspiredansatze1doi:10.1021/acs.jctc.8b01004}. However, this cannot be generalised to other areas such as designing variational error correction circuits or quantum optimisation problems. For example, in \cite{Xu2021-dt}, when developing a variational circuit that can encode logical states for the 5-qubit quantum error correction code, the authors adopted an expensive approach by randomly searching over a large number (order of 10000) of circuits. It is anticipated that, with the increasing number of application areas for VQCs and the need for scalability to tackle large problem sizes without relying on fundamental physical properties, such random search methods or methods based purely on human heuristics will struggle to find suitable ans\"atzes. Therefore, it is important to develop efficient methods for the automated design of variational quantum circuits. Here we focus on the development of algorithms for the automated design of VQCs by leveraging the power of artificial intelligence (AI) which can be deployed for a wide range of applications.


Although modern AI research often focuses on applications of image and natural language processing, the power of AI can also bring new knowledge in many areas, especially scientific discovery. 
AlphaFold2 managed to discover new mechanism for the bonding region of the protein and inhibitors \cite{alphafold2Jumper2021-lw} with competitive accuracy on predicting the three-dimensional structure of proteins in the 14th Critical Assessment of protein Structure Prediction (CASP) competition. In 2021, machine learning algorithms helped mathematicians discover new mathematical relationships in two different areas of mathematics \cite{Davies2021-xh}. Like variational quantum circuits, modern deep neural networks (DNN) also face a design problem when composing the network for certain tasks. With the help of AI algorithms, researchers developed techniques to efficiently search suitable network architectures in a large search space. Famous algorithms for neural architecture search (NAS) include the DARTS algorithm \cite{DARTS_DBLP:conf/iclr/LiuSY19}, which models the choice of operations placed in different layers as an independent categorical probabilistic model that can be optimised via gradient descent methods, and the PNAS algorithm \cite{PNAS10.1007/978-3-030-01246-5_2}, which models the search process with sequential model-based optimisation (SMBO) strategy. 
Tree-based algorithms were also proposed for NAS, such as AlphaX \cite{AlphaXDBLP:conf/aaai/WangZJTF20}, which models the search process similarly as the search stage of AlphaGo \cite{AlphaGoDBLP:journals/nature/SilverHMGSDSAPL16}. Recently, a new NAS algorithm based on tree search and combinatorial multi-armed bandits, proposed in \cite{huang2021neural}, outperforms other NAS algorithms, including the previously mentioned algorithms.

Based on progress in neural architecture search algorithms, efforts have been made on developing similar approaches for Quantum Ans\"atz (Architecture) Search (QAS) problems. Zhang \textit{et.al}~\cite{zhang2021differentiable} adapted the DARTS algorithm \cite{DARTS_DBLP:conf/iclr/LiuSY19} from NAS for QAS, which models the distribution of different operations within a single layer with the independent category probabilistic model. The search algorithm will update the parameters in the VQC as well as the probabilistic model. However, it has been shown in NAS literature that DARTS tend to assign fast-converge architectures with high probability during sampling \cite{Shu2019-jf, Zhou2020-dg}. Also, the off-the-shelf probabilistic distributions for modelling the architecture space tend to have difficulties when the search space is large. Later, the same group of authors developed a neural network to evaluate the performance of parameterised quantum circuits without actually training the circuits, and incorporated this neural network into quantum architecture search \cite{zhang2021neural}. While NAS algorithms often focus on image related tasks and it has been proved through many experiments that one neural network architecture can act as a backbone feature extractor for many downstream tasks, the structures of variational quantum circuits for different problems often vary a great deal with different problems, casting some doubts on the generalisation abilities of such neural predictor based QAS algorithms. Kuo \textit{et.al} \cite{kuo2021quantum} proposed a deep reinforcement learning based method for tackling QAS. The reinforcement learning agent is optimised by the advantage actor-critic and proximal policy optimisation algorithms. 
However, NAS algorithms based on policy gradient reinforcement learning have been shown to get easily stuck in local minimal, producing less optimal solutions \cite{ENASpmlr-v80-pham18a, Sutton1999-nj}. Also, the data size for training a reinforcement learning agent will explode when the number of actions the agent can choose from is large. He \textit{et.al} \cite{chen2021quantum} applied meta-learning techniques to learn good heuristics of both the architecture and the parameters. Du \textit{et al.}  \cite{du2020quantum} proposed a QAS algorithm based on the one-shot neural architecture search, where all possible quantum circuits are represented by a supernet with a weight-sharing strategy and the circuits are sampled uniformly during the training stage. After finishing the training stage, all circuits in the supernet are ranked and the best performed circuit will be chosen for further optimisation. Later Linghu \textit{et.al}~\cite{Linghu2022-yy} applied similar techniques on search to a classification circuit on a physical quantum processor. Meng \textit{et.al}~\cite{9566740mctsqas} applied Monte-Carlo tree search to ans\"atz optimisation for problems in quantum chemistry and condensed matter physics. However, these studies often restrict their demonstrations within one or two types of problems and small-sized systems.



In order to develop a search technique that can be applied to larger search spaces and different variational quantum problems, we introduce an algorithm for QAS problems based on combinatorial multi-armed bandit (CMAB) model as well as Monte-Carlo Tree Search (MCTS). In order to explore extremely large search spaces compared to previous work in the literature, the working of our strategy is underpinned by a reward scheme which dictates the choices of the quantum operations at each step of the algorithm with the na\"ive assumption \cite{CMAB_RTS}. This enabled our strategy to work on larger systems, more than 7 qubits, whereas the existing examples \cite{zhang2021differentiable, chen2021quantum, kuo2021quantum, zhang2021differentiable, du2020quantum, zhang2021neural} are restricted to typically 3 or 4 qubits, with the largest being 6 qubits. To demonstrate the working of our method, we showed its application to a variety of problems including encoding the logic states for the [[4,2,2]] quantum error detection code, solving the ground energy problem for different molecules as well as linear systems of equations, and searching the ans\"atz for solving optimisations problems. Our work confirms that the automated quantum architecture search based on the MCTS+CMAB approach exhibits great versatility and scalability, and therefore should provide an efficient solution and new insights to the problems of designing variational quantum circuits.

This paper is organised as follows: Section~\ref{methods} introduces the basic notion of Monte-Carlo tree search, as well as other techniques required for our algorithm, including nested MCTS and na\"ive assumptions from the CMAB model. Section \ref{experiments}  reports the results based on the application of our search algorithm to various problems, including searching for encoding circuits for the [[4,2,2]] quantum error detection code, the ans\"atz circuit for finding the ground state energy of different molecules, as well as circuits for solving linear system of equations and optimisation. In Section \ref{discussion} we discuss the results and conclusions.

\section{Methods}\label{methods}
\subsection{Problem Formulation}
In this paper, we formulate the quantum ans\"atz search problem, which is aimed to automatically design variational quantum circuits to perform various tasks, as a tree structure. We slice a quantum circuit into layers, and for each layer there is a pool of candidate operations. Starting with an empty circuit, we fill the layers with operations chosen by the search algorithm, from the first to the final layer. 
\begin{figure}[H]
  \centering
  \includegraphics[width=\textwidth]{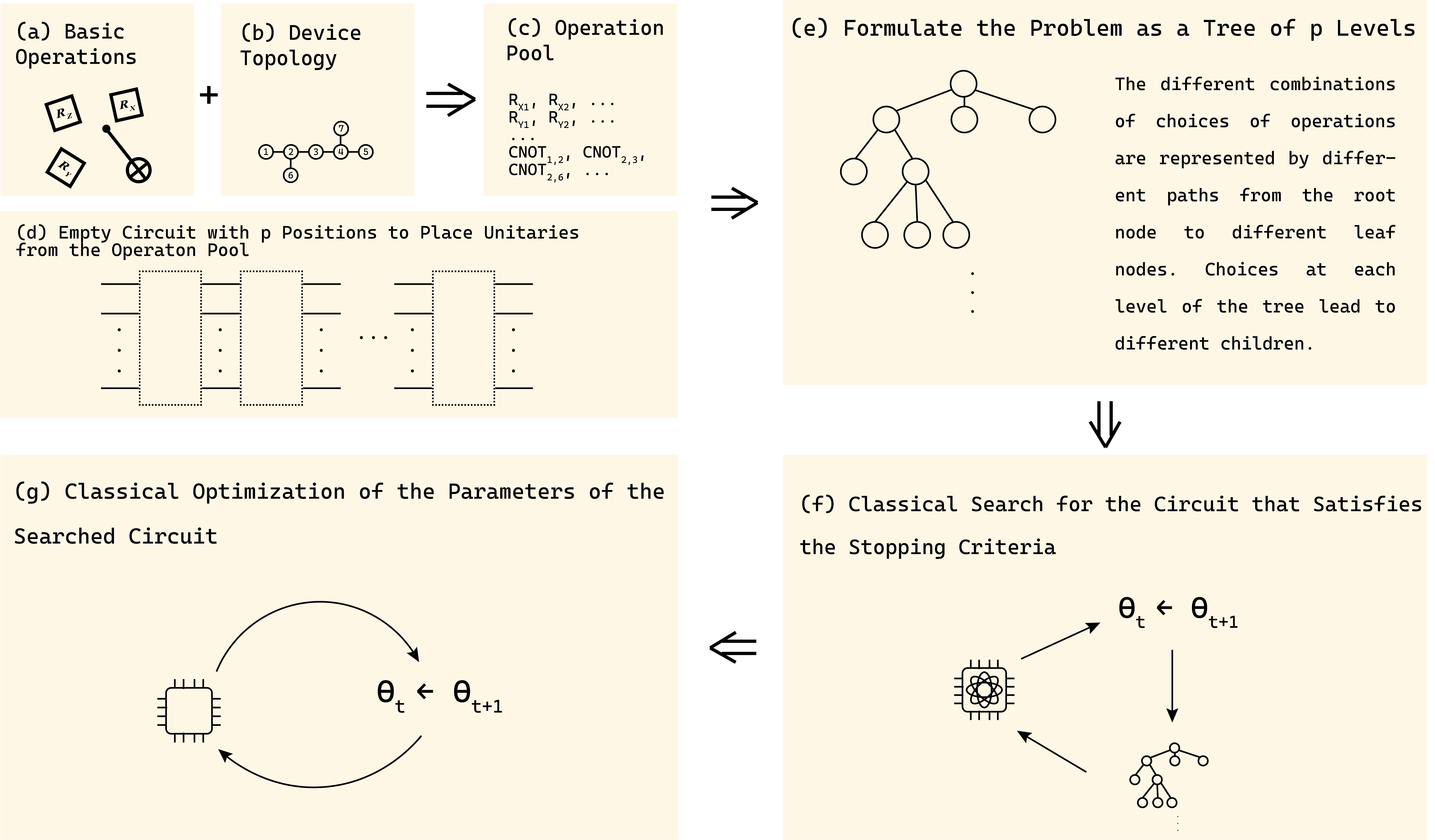}
  \caption{An overview of the algorithmic framework proposed in this paper. The operation pool (c) is obtained by tailoring the basic operations (a) with respect to the device topology (b). After that, we formulate the combinations of different choices of operations at different layer position in the circuit (d) as a search tree (e). In (f), we evaluate our circuit on a quantum processor or quantum simulator to get value of the loss or reward function, and according to the value of the loss/reward function we update the parameters on a classical computer, then use MCTS to search for the current best circuit. We then send the updated circuit structure together with the updated parameters to the quantum processor/simulator to obtain a new set of loss/reward values. The process depicted in (f) will repeat until a circuit that meets the stopping criteria is found. Then, as shown in (g), we will follow the usual process to optimize the parameters in the searched variational quantum circuit by classical-quantum hybrid computing.}
  \label{fig:overview}
\end{figure}

A quantum circuit is represented as a (ordered) list, $\mathcal{P}$, of operations of length $p$ chosen from the operation list. The length of this list is fixed within the problem.
The operation pool is a set 
\begin{equation}
\mathcal{C} = \{U_0, U_1, \cdots, U_{c-1} \},
\end{equation}
with $\vert \mathcal{C} \vert = c$ the number of elements. Each element $U_i$ is a possible choice for a certain layer of the quantum circuit. Such operations can be parameterised (e.g. the $R_Z(\theta)$ gate), or non-parameterised (e.g. the Pauli gates). A quantum circuit with four layers could, for instance, be represented as:
\begin{equation}
    \mathcal{P} = [U_0, U_1, U_2, U_1],
\end{equation}
where, according to the search algorithm, the operations chosen for the first, second, third and fourth layer are $U_0$, $U_1$, $U_2$, $U_1$. In this case, $p=4$ and the size of the operation pool $\vert \mathcal{C} \vert = c$. The search tree is shown in Fig. \ref{fig:treeexample} In this paper, we will only deal with unitary operations or unitary channels. The output state of such a quantum circuit can then be written as:
\begin{equation}
    \vert\varphi_{\rm out}\rangle = U_1 U_2 U_1 U_0 \vert\varphi_{\rm init}\rangle\label{eq:U1U3U1U2},
\end{equation}
where $\vert \varphi_{\rm init}\rangle$ is the initial state of the quantum circuit. For simplicity, we will use integers to denote the chosen operations (such operations can be whole-layer unitaries, like the mixing Hamiltonians often seen in typical QAOA circuits, or just single- and two-qubit gates). 
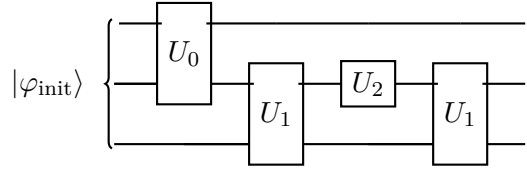
\begin{figure}[H]
  \centering
  \begin{quantikz}[transparent, row sep={0.8cm,between origins}]
\qw & \midstick[wires=3,brackets=right]{$|\varphi_{\rm init}\rangle$} & \gate[2,disable auto height]{U_0} & \qw & \qw & \qw & \qw\\
\qw &  & \qw & \gate[2,disable auto height]{U_1} & \gate{U_2} & \gate[2,disable auto height]{U_1} & \qw\\
\qw &  & \qw & \qw & \qw & \qw & \qw
\end{quantikz}
  \caption{An example of the circuit corresponding to the series of unitaries applied to $\vert \varphi_{\rm init}\rangle$ in Eqn.\ref{eq:U1U3U1U2}.}
  \label{fig:U1U3U1U2_circ}
\end{figure}
For example, the  quantum circuit from Eqn.~\ref{eq:U1U3U1U2} can be written as:
\begin{equation}
    \mathcal{P} = [0, 1, 2, 1]\label{eq:U1U3U1U2_list}
\end{equation}
and the operation at the $i^{th}$ layer can be referred as $k_i$. For example, in the quantum circuit above, we have $k_2=1$.

\begin{figure}[H]
  \centering
  \includegraphics[width=0.4\textwidth]{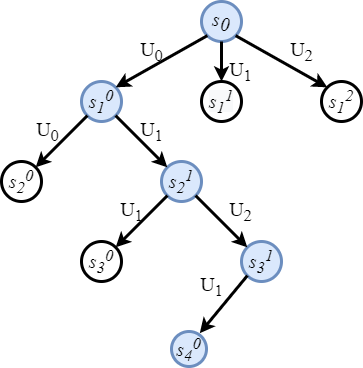}
  \caption{The tree representation (along the arc with blue-shaded circles) of the unitary described in Eqns. \ref{eq:U1U3U1U2} and \ref{eq:U1U3U1U2_list} as well as Fig.~\ref{fig:U1U3U1U2_circ}. The circle with $s_0$ is the root of the tree, which represents an empty circuit. Other circles with $s_i^j$ in it denote the $j^{th}$ node at the $i^{th}$ level of the tree. $i$ can also indicate the number of layers currently in the circuit at state $s_i^j$. For example, on the leftmost branch of the tree, there is a node labelled $s_2^0$, indicating that it is the $0^{th}$ node at level $2$. At $s_2^0$, the circuit would be $\mathcal{P}_{s_2^0}=[U_0, U_0]$, which clearly only has 2 layers. We can also see that some of the possible branches along the blue-node path are pruned, leading to the size of operation pool at some node smaller than the total number of possible choices $c = \vert \mathcal{C}\vert$.}
  \label{fig:treeexample}
\end{figure}

The performance of the quantum circuit can be evaluated from the loss $\mathcal{L}$ or reward $\mathcal{R}$, where the reward is just the negative of the loss. Both are functions of $\mathcal{P}$, and the parameters of the chosen operations $\boldsymbol{\theta}$:
\begin{equation}
    \mathcal{L}(\mathcal{P},\boldsymbol{\theta})=L(\mathcal{P}, \boldsymbol{\theta})+\lambda
\end{equation}

\begin{equation}
    \mathcal{R}(\mathcal{P},\boldsymbol{\theta})=R(\mathcal{P}, \boldsymbol{\theta})-\lambda,
\end{equation}
where $\lambda$ is some penalty function that may only appear when certain circuit structures appear, as well as other kinds of penalty terms, like penalty on the sum of absolute value of weights or the number of certain type of gates in the circuit; $L$ and $R$ are the loss/reward before applying the penalty. The purpose of the penalty term $\lambda$ is to `sway' the search algorithm from structures we do not desire.
Instead of storing all the operation parameters for each different quantum circuit, we share the parameters for a single operation at a certain location. That is, we have a multidimensional array of shape $(p, c, l)$, where $l$ is the maximum number of parameters for the operations in the operation pool. If all the operations in the pool are just the $U3$ gate \cite{nielsen00}:
\begin{equation}
U 3(\theta, \phi, \lambda)=\left[\begin{array}{cc}
\cos \left(\frac{\theta}{2}\right) & -e^{i \lambda} \sin \left(\frac{\theta}{2}\right) \\
e^{i \phi} \sin \left(\frac{\theta}{2}\right) & e^{i(\phi+\lambda)} \cos \left(\frac{\theta}{2}\right)
\end{array}\right]
\end{equation}
as well as its controlled version $CU3$ gate on different (pairs of) qubits, then in this case $l=3$.

To reduce the space required to store the parameters of all possible quantum circuits, for a quantum circuit with operation $k$ at layer $i$, the parameter is the same at that layer for that specific operation is the same for all other circuits with the same operation at the same location, which means we are sharing the parameters of the unitaries in the operation pool with other circuits. For example, in Fig.\ref{fig:treeexample}, besides the blue-node arc $\mathcal{P}=[U_0, U_1, U_2, U_1]$, there are also other paths, such $\mathcal{P}^{'} = [U_0, U_1, U_1, \cdots]$, and since the first two operations in $\mathcal{P}$ and $\mathcal{P}^{'}$ are the same, then we will share the parameters of $U_0$ and $U_1$ between these two circuits by setting the parameters to be the same for the $U_0$ and $U_1$ in both circuits, respectively.
Such a strategy is often called ``parameter-sharing'' or ``weight-sharing'' in the neural architecture search literature.

As shown in Fig~\ref{fig:treeexample} and mentioned earlier, the process of composing or searching a circuit can be formulated in the form of the tree structure. For example, if we start from an empty list $P = [\;]$ with maximal length four and an operation pool with three elements $C = \{U_0, U_1, U_2\}$, 
then the state of the root node of our search tree will be the empty list $s_0^0 = [\;]$. The root node will have three possible actions (if there are no restrictions on what kind of operations can be chosen), which will lead us to three children nodes with states $s_1^0 = [U_0]=[0], s_1^1 = [U_1]=[1], s_1^2=[U_2] = [2]$. For each of these nodes, there will be a certain number of different operations that can be chosen to append the end of the list, depending on the specific restrictions. There will always be a ``placeholder'' operation that can be chosen if all other operations fail to meet the restrictions. The penalty resulting from the number of ``placeholder'' operations will only be reflected in the loss (or reward) of the circuit. The nodes can always be expanded with different actions, leading to different children, until the maximum length of the quantum circuit has been reached, which will give us the leaf node of the search tree. 

The process of choosing operations at each layer can be viewed as a both a \textit{local} and \textit{global} multi-armed bandit (MAB). A multi-armed bandit, just as its name indicates, is similar to a bandit, or slot machine (in the casino), but has multiple levers, or arms, that can be pulled. Or equivalently, it can be viewed as someone who has multiple arms (maybe Squidward) that can pull the levers on different slot machines. In both cases, the rewards obtained from pulling different arms follow different (often unknown) distributions. The person pulling these arms needs to develop a strategy that can maximise his rewards from the machine(s). If we consider the whole circuit search problem as an MAB (the global MAB, $MAB_g$), then the "arms" are different circuit configurations. Although the rewards of these circuits are relatively easy to obtain based on the value of their cost functions after training of the circuits is finished (which still requires a fair amount of time for training), the exploding number of possible circuit configurations when the size of operation pool and number of layers increase makes it impossible to perform an informed search for suitable solutions while training every circuit we encountered during the search process. Since our circuit is basically a combination of different choice of layer unitaries, we can decompose the whole problem into the choices of unitaries at each layer, which is the local MAB, $MAB_i$, $i$ denoting the MAB problem from choosing the suitable unitary at layer $i$. In the local MAB for a single layer, the "arms" of the MAB are no longer the circuit configuration, instead the (permitted) unitary operations from the operation pool $\mathcal{C}$. Although the number of choices for the local MABs is considerably smaller than the global MAB, the reward for each arm is not directly observable. In next section, we will introduce the na\"ive assumption \cite{CMAB_RTS} to approximate the rewards of the local MABs from the global MAB, which will help us determine the rewards of the actions on each node (state) on the search tree for MCTS.

\begin{itemize}
    \item \textit{Local MAB}: The choice of unitary operations at each layer can be considered a \textit{local} MAB. That is, different unitary operations can be treated as different ``arms'' of the bandit;
    \item \textit{Global MAB}: We can also treat the composition of the entire quantum circuit as a \textit{global} MAB. That is, different quantum circuits can be viewed as different ``arms'' of the global bandit.
\end{itemize}


\subsection{Monte Carlo tree search (MCTS), nested MCTS and the na\"ive assumption}
Monte Carlo tree search (MCTS) is a heuristic search algorithm for a sequence decision process. It has achieved great success in other areas, including defeating the 18-time world champion Lee Sedol in the game of Go \cite{AlphaGoDBLP:journals/nature/SilverHMGSDSAPL16, AlphaGoZeroDBLP:journals/nature/SilverSSAHGHBLB17}. Generally, there are four stages in a single iteration of MCTS (see Fig.~\ref{fig:mcts}) \cite{MCTS_for_game10.5555/3022539.3022579}:
\begin{figure}[]
  \centering
  \includegraphics[width=0.8\textwidth]{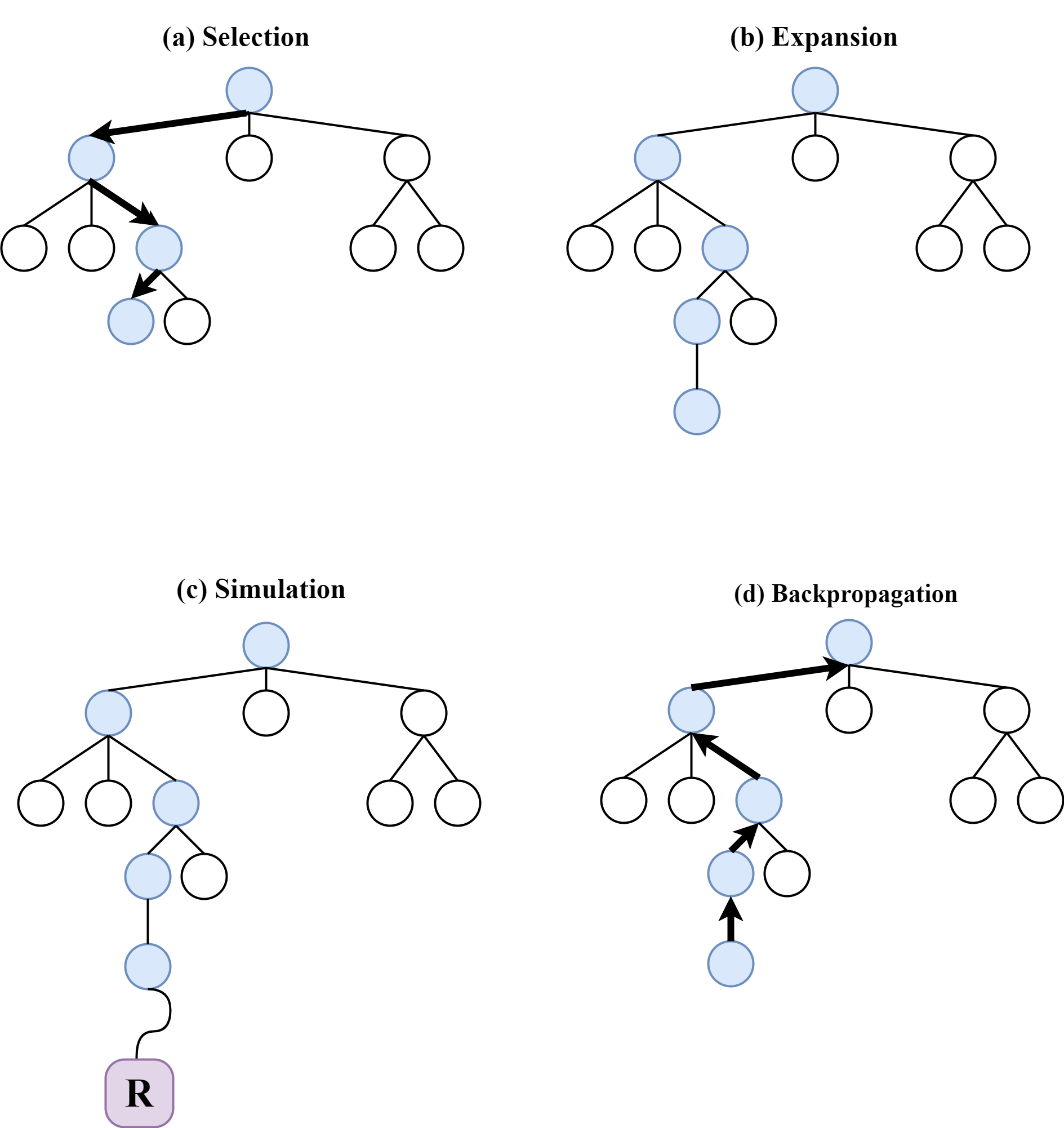}
  \caption{Four stages of Monte Carlo tree search. From left to right, up to down: Selection: Go down from the root node to a non fully expanded leaf node; Expansion: Expand the selected node by taking an action; Simulation: Simulate the game, which in our case is the quantum circuit, to obtain reward information \textbf{R}; Backpropagation: Back-propagation of the reward information along the path (arc) taken.}
  \label{fig:mcts}
\end{figure}

\begin{itemize}
    \item \textsc{Selection:}(Fig.\ref{fig:mcts}(a)) In the selection stage, the algorithm will, starting from the root of the tree, find a node at the end of an arc (a path from the root of the tree to the leaf node, the path marked by bold arrows and blue circles in Fig.\ref{fig:mcts}). The nodes along the arc are selected according to some policy, often referred as the ``selection policy'', until a non fully expanded node or a leaf node is reached. If the node is a leaf node, i.e after selecting the operation for the last layer of the quantum circuit, we can directly jump to the simulation stage to get the reward of the corresponding arc. If the node is not a leaf node, i.e the node is not fully expanded, then we can progress to the next stage;
    \item \textsc{Expansion:}(Fig.\ref{fig:mcts}(b)) In the expansion stage, at the node selected in the previous stage, we choose a previously unvisited child by choosing a previously unperformed action. We can see from the upper right tree in Fig.\ref{fig:mcts} that a new node has been expanded at the end of the arc; 
    \item \textsc{Simulation:}(Fig.\ref{fig:mcts}(c)) In the simulation stage, if the node obtained from the previous stages is not a leaf node, we continue down the tree until we have reached a leaf node, i.e finish choosing the operation for the last layer. After we have the leaf node, we simulate the circuit and obtain the loss $\mathcal{L}$ (or reward $\mathcal{R}$). Usually, the loss $\mathcal{L}$ is required to update the parameters in the circuit;
    \item \textsc{Backpropagation:}(Fig.\ref{fig:mcts}(d)) In this stage, the reward information obtained from the simulation stage is back-propagated through the arc leading from the root of the tree to the leaf node, and the number of visits as well as the (average) reward for each node along the arc is be updated.
\end{itemize}

\begin{figure}[H]
  \centering
  \includegraphics[width=0.8\textwidth]{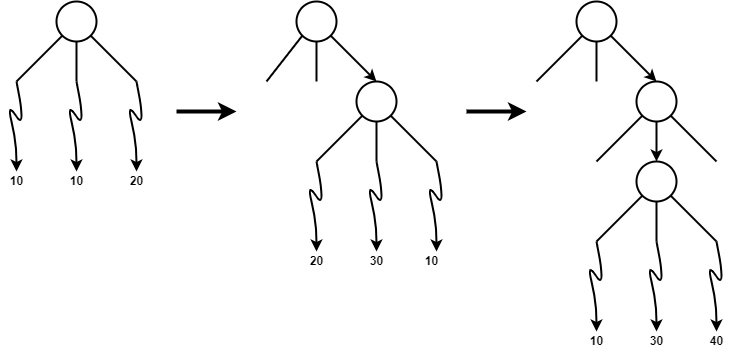}
  \caption{Nested Monte Carlo tree search. Left: The root node has three possible actions, which in this case are unselected initially. We perform MCTS on all three children nodes (generated by the three possible actions) to update their reward information. After one iteration of MCTS with each child as root node for the search tree that MCTS performed on, the rewards of these three actions leading to the three child nodes are 10, 10, 20, respectively. In this case, the right child has the highest reward. Middle: After selecting the right side child node, we perform the same MCTS on all three possible children nodes as before, which gives updated reward information. In this case the middle child node has the highest reward, meaning that at this level we expand the middle child node. Right: Similar operations as before. If we only perform nested MCTS at the root node level, then it will be a level-1 nested MCTS.}
  \label{fig:nestedmcts}
\end{figure}

The nested MCTS algorithm \cite{nestedmontecarlosearch} is based on the vanilla MCTS algorithm. However, before selecting the best child according to the selection policy, a nested MCTS will be performed on the sub-trees with each child as the root node. Then the best child will be selected according to the selection policy with updated reward information, see Fig.~\ref{fig:nestedmcts}.

We denote a quantum circuit with $p$ layers $\mathcal{P} = [k_1,\cdots, k_p]$, with each layer $k_i$ having a search space no greater than $\vert \mathcal{C} \vert = c$ (where $c$ is the number of possible unitary operations, as defined earlier). Then each choice for layer $k_i$ is a \textit{local arm} for the \textit{local MAB}, $MAB_i$. The set of these choices is also denoted as $k_i$. The combination of all $p$ layers in $\mathcal{P}$ forms a valid quantum circuit, which is called a \textit{global arm} of the \textit{global MAB, $MAB_g$}. 

Since the global arm can be formed from the combination of the local arms, if we use the na\"ive assumption \cite{CMAB_RTS}, the global reward $R_{\rm global}$ for $MAB_g$ can be approximated by the sum of the reward of local MABs, and each local reward only depends on the choice made in each local MAB. This also means that, if the global reward is more easily accessed than the local rewards, then the local rewards can be approximated from the global reward. With the na\"ive assumption, we can have a linear relationship between the global reward and local rewards:
\begin{equation}
    R_{\rm global} = \frac{1}{p}\sum_{i=1}^p R_i
\end{equation}
When searching for quantum circuits, we have no access to the reward distribution of individual unitary operations, however, we can apply the na\"ive assumption to approximate those rewards (``local reward'') with the global reward: 
\begin{equation}
    R_{i} \approx R_{\rm global}
\end{equation}
where $R_{i}$ is the reward for pulling an arm at \textit{local $MAB_i$} and $R_{\rm global}$ is the reward for the global arm.
Also, if we use the na\"ive assumption, we will not need to directly optimise on the large space of global arms as in  traditional MABs. Instead, we can apply MCTS on the local MABs to find the best combination of local arms.

In the original work on nested MCTS~\cite{nestedmontecarlosearch}, a random policy was adopted for sampling. In this paper we will instead change it to the famous UCB policy~\cite{UCB_paper_10.5555/944919.944941}. Given a local $MAB_i$, with the set of all the possible choices $k_i$, the UCB policy can be defined as:
\begin{equation}
    UCB: \argmax_{arm_j\in k_i} \Bar{R}(k_i, arm_j) + \alpha \sqrt{\frac{2\ln n_i}{n_j}}
\end{equation}
where $\Bar{R}(k_i, arm_j)$ is the average reward for $arm_j$ (i.e the reward for operation choice $U_j$  for layer $k_i$) in local $MAB_i$, $n_i$ is the number of times that $MAB_i$ has been used and $n_j$ is the number of times that $arm_j$ has been pulled. The parameter $\alpha$ provides a balance between exploration ($\sqrt{\frac{2\ln n_i}{n_j}}$) and exploitation ($\Bar{R}(k_i, arm_j$)). The UCB policy modifies the reward which the selection of action will be based on.

For small $\alpha$, the actual reward from the bandit will play a more important role in the UCB modified rewards, which will lead to selecting actions with previously observed high rewards. When $\alpha$ is large enough, the second term, which will be relatively large if $MAB_i$ has been visited many times but $arm_j$ of $MAB_i$ has only been pulled a small number of times, will have more impact on the modified reward, leading to a selection favoring previously less visited actions.

\subsection{QAS with Nested Na\"ive MCTS}


    
Generally, a single iteration for the search algorithm will include two steps for non-parameterised circuits, and two more parameter-related steps for parameterised quantum circuits. The set of parameters, which will be referred to as the parameters of the super circuit, or just parameters, in the following algorithms, follow the same parameter sharing strategy as described in Section 2.1. That is, if the same unitary operation (say, $U_2$) appears in the same location (say, layer \#5) across different quantum circuits, then the parameters are the same, even for different circuits. Also, with parameterised quantum circuits (PQC), it is common practice to ``warm-up'' the parameters by randomly sampling a batch of quantum circuits, calculating the averaged gradient, and update the parameters according to the averaged gradient, to get a better start for the parameters during the search process. During one iteration of the search algorithm, we have:
\begin{enumerate}
    \item Sample a batch of quantum circuits from the super circuit with Algorithm \ref{alg:sampleArc};
    \item (For PQCs) Calculate the averaged gradients of the sampled batch, add noise to the gradient to guide the optimiser to a more ``flat'' minimum if needed;
    \item (For PQCs) Update the super circuit parameters according to the averaged gradients;
    \item Find the best circuit with Algorithm \ref{alg:exploitArc}.
\end{enumerate}

We could also set up an early-stopping criteria for the search. That is, when the reward of the circuit obtained with Algorithm \ref{alg:exploitArc} meets a pre-set standard, we will stop the search algorithm and return the circuit that meet such standard (and further fine-tune the circuit parameters if there are any).

With the na\"ive assumption, which means the reward is evenly distributed on the local arms pulled for a global MAB, we can impose a prune ratio during the search. That is, given a node that has child nodes, if the average reward of a child node is smaller than a ratio, or percentage, of the average reward of the said node, then this child node will be removed from the set of all children, unless the number of children reached the minimum requirement.

\begin{algorithm}
\caption{SampleArc}\label{alg:sampleArc}
\begin{algorithmic}
\Require sample policy $Policy$, parameters of the super circuit $param$, number of rounds in sampling $N$
\Ensure list representation $\mathcal{P}$ of quantum circuit
\State $curr \gets GetRoot(Tr)$ \Comment{Starting from the root node of the tree $Tr$}
\State $i\gets0$ \Comment{Counter}
\While{$i<N$}
\State $ExecuteSingleRound(curr, Policy, param)$
\State $i\gets i+1$
\EndWhile
\While{$curr$ is not leaf node}
\State $curr\gets SelectNode(curr, Policy)$
\EndWhile
\State $\mathcal{P}\gets GetListRepresentation(curr)$
\end{algorithmic}
\end{algorithm}

\begin{algorithm}[H]
\caption{ExploitArc}\label{alg:exploitArc}
\begin{algorithmic}
\Require exploit policy $Policy$, parameters of the super circuit $param$, number of rounds in exploitation $N$
\Ensure list representation $\mathcal{P}$ of quantum circuit
\State $curr \gets GetRoot(Tr)$ \Comment{Starting from the root node of the tree $Tr$}
\While{$curr$ is not leaf node}
\State $i\gets0$ \Comment{Counter}
\While{$i<N$}
\State $ExecuteSingleRound(curr, Policy, param)$
\State $i\gets i+1$
\EndWhile
\State $curr\gets SelectNode(curr, Policy)$
\EndWhile
\State $\mathcal{P}\gets GetListRepresentation(curr)$
\end{algorithmic}
\end{algorithm}

\begin{algorithm}
\caption{SelectNode}\label{alg:selectChild}
\begin{algorithmic}
\Require current node $n$, selection policy $Policy$
\Ensure selected node $n'$

\If{$n$ is fully expanded}
    \State $PruneChild(n)$ \Comment{Prune children nodes according to certain threshold}
    \State $n' \gets GetBestChild(n, Policy)$  \Comment{Select the best child}
\Else
    \State $n' \gets ExpandChild(n)$ \Comment{Expand the node}
\EndIf
\end{algorithmic}
\end{algorithm}

\begin{algorithm}[H]
\caption{ExecuteSingleRound}\label{alg:executeSingleRound}
\begin{algorithmic}
\Require current node $n$, selection policy $Policy$, parameters of the super circuit $param$
\Ensure leaf node $n'$
\State $n'\gets n$
\While{$n'$ is not leaf node}
\State $n'\gets SelectNode(n', Policy)$
\EndWhile
\State $R \gets Simulation(n', param)$  \Comment{Obtain reward from simulation}
\State $Backpropagate(n', R)$ \Comment{Back-propagate the reward information along the arc}
\end{algorithmic}
\end{algorithm}

\section{Numerical Experiments and Results}\label{experiments}
\subsection{Searching for the encoding circuit of [[4,2,2]] quantum error detection code}\label{422}

The [[4,2,2]] quantum error detection code is a simple quantum error detection code, which needs 4 physical qubits for 2 logical qubits and has a code distance 2. It is the smallest stabilizer code that can detect X- and Z-errors \cite{qec_intro_guide}. One possible set of code words for the [[4,2,2]] error detection code is:

\begin{equation}
\mathcal{E}_{\rm [[4,2,2]]}=\operatorname{span}\left\{\begin{array}{l}
|00\rangle_{L}=\frac{1}{\sqrt{2}}(|0000\rangle+|1111\rangle) \\
|01\rangle_{L}=\frac{1}{\sqrt{2}}(|0110\rangle+|1001\rangle) \\
|10\rangle_{L}=\frac{1}{\sqrt{2}}(|1010\rangle+|0101\rangle) \\
|11\rangle_{L}=\frac{1}{\sqrt{2}}(|1100\rangle+|0011\rangle)
\end{array}\right\}
\end{equation}

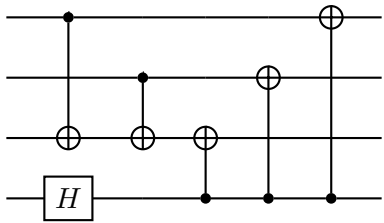
\begin{figure}[H]
  \centering
  \begin{quantikz}[transparent, row sep={0.8cm,between origins}]
\qw & \ctrl{0} & \qw & \qw & \qw & \targ{}\vqw{0} & \qw\\
\qw & \qw & \ctrl{0} & \qw & \targ{}\vqw{0} & \qw & \qw\\
\qw & \targ{}\vqw{-2} & \targ{}\vqw{-1} & \targ{}\vqw{0} & \qw & \qw & \qw\\
\qw & \gate{H} & \qw & \ctrl{-1} & \ctrl{-2} & \ctrl{-3} & \qw
\end{quantikz}
  \caption{Encoding circuit of the [[4,2,2]] code \cite{qec_intro_guide} to detect X- and Z-errors. It needs 4 physical qubits for 2 logical qubits and has a code distance 2. By our settings, the number of layers equals to the number of operations in the circuit. In this figure, the number of layers is 6.}
\label{fig:lit422}
\end{figure}

The corresponding encoding circuit is shown in Fig.\ref{fig:lit422}.

Quantum error detection and correction is vital to large-scale fault-tolerant quantum computing. By searching for the encoding circuit of the [[4,2,2]] error detection code, we demonstrate that our algorithm has the potential to automatically find  device-specific encoding circuits of quantum error detection and correction codes for future quantum processors.

\subsubsection{Experiment Settings}
When searching for the encoding circuit of the [[4,2,2]] quantum error correction code, we adopted an operation pool consisting of only non-parametric operations: the Hadamard gate on each of the four qubits and CNOT gates between any two qubits. The total size of the operation pool is $4 + \frac{4!}{2!\times2!}\times2=16$. When there are 6 layers in total, the overall size of the search space is $16^6\approx1.67\times10^7$.

The loss function for this task is based on the fidelity between the output state of the searched circuit and the output generated by the encoding circuit from Section 4.3 of \cite{qec_intro_guide} (also shown in Fig.~\ref{fig:lit422}) when input states taken from the set of Pauli operator eigenstates and the magic state $\vert T \rangle$ are used:
\begin{equation}
    \mathcal{S}=\{\vert 0 \rangle, \vert 1 \rangle, \vert + \rangle, \vert - \rangle, \vert +i \rangle, \vert -i \rangle, \vert T \rangle\}
\end{equation}
where $\vert T \rangle = \frac{\vert 0 \rangle + e^{i\pi/4}\vert 1 \rangle}{\sqrt{2}}$.


The input states (initialised on all four qubits) are
\begin{equation}
    \mathcal{I}_{[[4,2,2]]} = \{\vert \varphi_1 \rangle \otimes \vert \varphi_2 \rangle\otimes\vert 00\rangle \; \vert \; \vert \varphi_1 \rangle, \vert \varphi_2 \rangle \in \mathcal{S}\}
\end{equation}

We denote the unitary on all four qubits shown in Fig.~\ref{fig:lit422} as $U_{[[4,2,2]]}$, and the unitary from the searched circuit as $U_{Searched\;[[4,2,2]]}$, which is a function of the structure $\mathcal{P}_{Searched\;[[4,2,2]]}$. The loss and reward function can then be expressed as:
\begin{equation}
    L_{[[4,2,2]]} = 1-\frac{1}{\vert \mathcal{I}_{[[4,2,2]]} \vert}\sum_{\vert \psi_i\rangle\in  \mathcal{I}_{[[4,2,2]]}} \langle \psi_i \vert U_{\rm Searched\;[[4,2,2]]}^{\dagger} O_{[[4,2,2]]}(\vert \psi_i\rangle)  U_{\rm Searched\;[[4,2,2]]} \vert \psi_i\rangle
\end{equation}
\begin{equation}
    R_{[[4,2,2]]} = 1-L_{[[4,2,2]]}
\end{equation}
where
\begin{equation}
    O_{[[4,2,2]]}(\vert \psi_i\rangle) = U_{[[4,2,2]]} \vert \psi_i\rangle \langle \psi_i \vert U_{[[4,2,2]]}^{\dagger},\; \vert \psi_i\rangle\in  \mathcal{I}_{[[4,2,2]]}
\end{equation}

The circuit simulator used in this and the following numerical experiments is Pennylane \cite{bergholm2020pennylane}.
\subsubsection{Results}
\begin{figure}[H]
    \centering
    \includegraphics[width=0.8\textwidth]{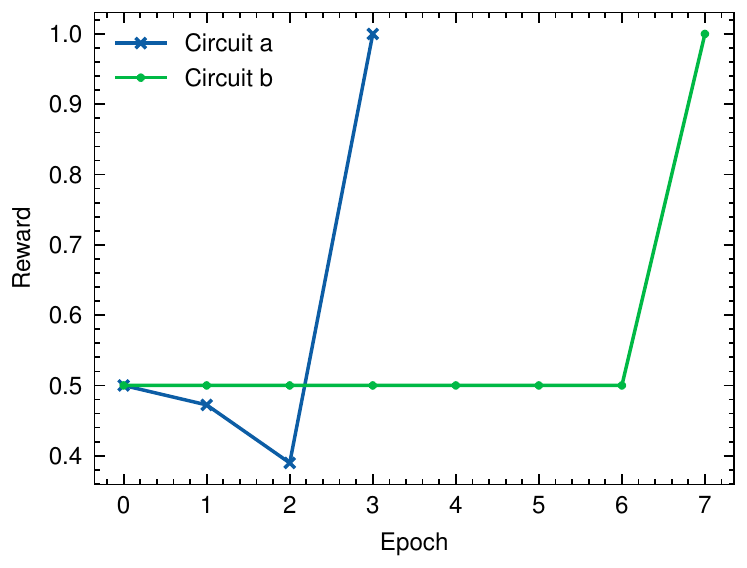}
    \caption{Rewards when searching for encoding circuits of the [[4,2,2]] code. We can see that in both cases the algorithm was able find the encoding circuit that generated the required code words in just a few iterations. `Circuit a' refers to the search rewards for the circuit in Fig.\ref{fig:422_first_circ} and `Circuit b' refers to the search rewards for the circuit in Fig.\ref{fig:422_second_circ}.}\label{fig:422_reward}
\end{figure}

\begin{figure}[H]
    \centering
    \begin{subfigure}[b]{0.48\textwidth}
    \centering
        \begin{quantikz}[transparent, row sep={0.8cm,between origins}]
\qw & \qw & \ctrl{0} & \targ{}\vqw{0} & \qw & \qw & \ctrl{0} & \qw\\
\qw & \qw & \qw & \qw & \ctrl{0} & \targ{}\vqw{0} & \qw & \qw\\
\qw & \qw & \qw & \qw & \targ{}\vqw{-1} & \qw & \targ{}\vqw{-2} & \qw\\
\qw & \gate{H} & \targ{}\vqw{-3} & \ctrl{-3} & \qw & \ctrl{-2} & \qw & \qw
\end{quantikz}
        \caption{}
        \label{fig:422_first_circ}
    \end{subfigure}
    ~ 
    \begin{subfigure}[b]{0.48\textwidth}
    \centering
        \begin{quantikz}[transparent, row sep={0.8cm,between origins}]
\qw & \ctrl{0} & \qw & \qw & \ctrl{0} & \targ{}\vqw{0} & \qw\\
\qw & \targ{}\vqw{-1} & \targ{}\vqw{0} & \ctrl{0} & \targ{}\vqw{-1} & \qw & \qw\\
\qw & \qw & \qw & \targ{}\vqw{-1} & \qw & \qw & \qw\\
\qw & \gate{H} & \ctrl{-2} & \qw & \qw & \ctrl{-3} & \qw
\end{quantikz}

        \caption{}
        \label{fig:422_second_circ}
    \end{subfigure}
    \caption{Two different encoding circuits of the [[4,2,2]] code produced by the search algorithm.}\label{fig:422_circ}
\end{figure}
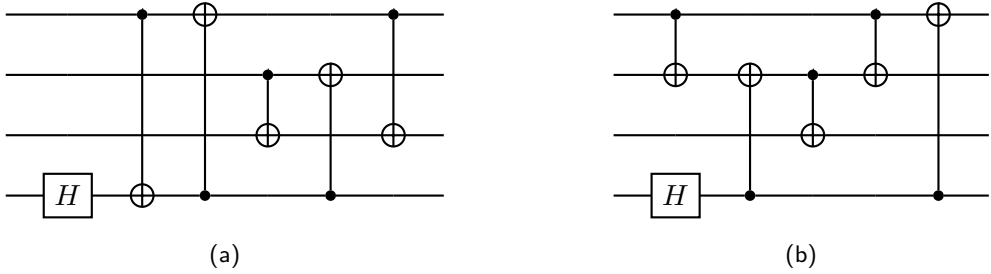

To verify whether the search algorithm will always reach the same solution, we ran the search algorithm twice, and both times the algorithm found an encoding circuit within a small numbers of iterations (Fig.~\ref{fig:422_reward}), although the actual circuit are different from each other, as shown in Fig.~\ref{fig:422_circ}. The search process that gave the circuit in Fig.\ref{fig:422_first_circ} met the early-stopping criteria in four iterations, and the search process that gave the circuit in Fig.\ref{fig:422_second_circ} met the early-stopping criteria in eight iterations, as shown in Fig.~\ref{fig:422_reward}.

\subsection{Solving linear equations}
The variational quantum linear solver (VQLS), first proposed in \cite{Bravo-Prieto_undated-oq}, is designed to solve linear systems $Ax=b$ on near term quantum devices. Instead of using quantum phase estimation like the HHL algorithm \cite{HHL}, which is unfeasible on near term devices due to large circuit depth, VQLS adopts a variational circuit to prepare a state $\ket{x}$ such that 
\begin{equation}
    A\ket{x} \propto \ket{b}
\end{equation}
In this section, we will task our algorithm to automatically search for a variantional circuit to prepare a state $\ket{x}$ to solve $Ax = b$ with A in the form of 
\begin{equation}
    A = \sum_l c_l A_l
\end{equation}
where $A_l$ are unitaries, and $\ket{b} = H^{\otimes n}\ket{\mathbf{0}}$.
\noindent
We will also adopt the local cost function $C_L$ described in \cite{Bravo-Prieto_undated-oq}:
\begin{equation}\label{eqn:vqls_local_loss}
C_{L}=1-\frac{\sum_{l, l^{\prime}} c_{l} c_{l^{\prime}}^{*}\bra{0}V^{\dagger} A_{l^{\prime}}^{\dagger} U P U^{\dagger} A_{l} V\ket{0}}{\sum_{l, l^{\prime}} c_{l} c_{l^{\prime}}^{*}\bra{0}V^{\dagger} A_{l^{\prime}}^{\dagger} A_{l} V\ket{0}}
\end{equation}
where $U=H^{\otimes n}$, $V$ is the (searched) variational circuit that can produce the solution state $V\ket{0} = \ket{x}$, and $P=\frac{1}{2}+\frac{1}{2 n} \sum_{j=0}^{n-1} Z_{j}$ \cite{pennylane_vqls}.

\subsubsection{Experiment Settings}
The linear system to be solved in our demonstration is:
\begin{equation}
    A = \zeta  I + J X_1 + J X_2  + \eta Z_3 Z_4
\end{equation}
\begin{equation}
    \ket{b} = H^{\otimes 4}\ket{0}
\end{equation}
with $J = 0.1, \zeta = 1, \eta = 0.2$.
The loss function we adopted follows the local loss $C_L$ in Eqn.~\ref{eqn:vqls_local_loss}. However, since the starting point of the loss values often has a magnitude of $10^{-2}\sim 10^{-3}$, we will need scaling in the reward function:
\begin{equation}
    \mathcal{R} = e^{-10 C_L}-\lambda
\end{equation}
where $\lambda$ is a penalty term depending on the number of Placeholder gates in the circuit. The operation pool consists of CNOT gates between neighbouring two qubits as well as the first and fourth qubits, the Placeholder and the single qubit rotation gate Rot \cite{nielsen00}:
\begin{equation}
Rot(\phi, \theta, \omega)=R_Z(\omega) R_Y(\theta) R_Z(\phi)=\left[\begin{array}{cc}
e^{-i(\phi+\omega) / 2} \cos (\theta / 2) & -e^{i(\phi-\omega) / 2} \sin (\theta / 2) \\
e^{-i(\phi-\omega) / 2} \sin (\theta / 2) & e^{i(\phi+\omega) / 2} \cos (\theta / 2)
\end{array}\right]
\end{equation}
The size of the operation pool $c = \vert \mathcal{C}\vert = 16$, and number of layers $p = 10$, giving us a search space of size $\vert \mathcal{S} \vert = 10^{16}$. There is also an additional restriction of maximum number of CNOT gates in the circuit, which is 8, the number of CNOT gates required to created two layers of circular entanglement.

\subsubsection{Results}
\begin{figure}[H]
    \centering
    \begin{subfigure}[t]{0.48\textwidth}
        \includegraphics[width=\textwidth]{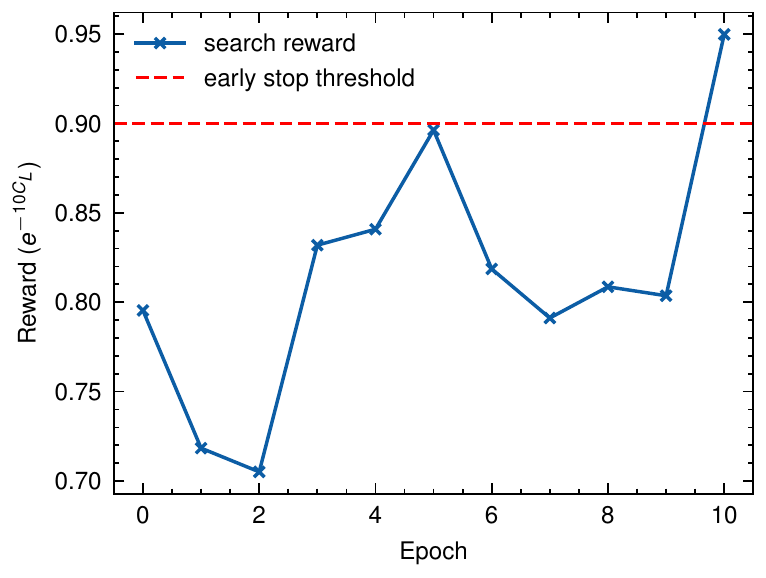}
        \caption{Search rewards for VQLS. The change of rewards with respect to the iterations is shown. We can see that the reward quickly reached the early stopping threshold at iteration 10. In the VQLS case, the reward is scaled since the initial reward with random sampled circuit structure and parameters is already at the magnitude of $10^{-2}$.}
        \label{fig:vqls_4q_search}
    \end{subfigure}
    ~ 
    \begin{subfigure}[t]{0.48\textwidth}
        \includegraphics[width=\textwidth]{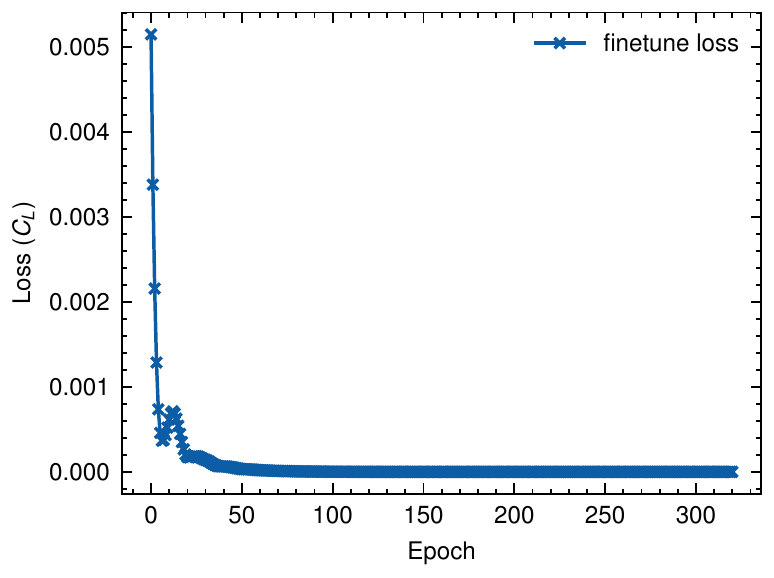}
        \caption{Fine-tune loss for the VQLS circuit.After the searched stopped at iteration 10 as shown in Fig~\ref{fig:vqls_4q_search}, the structure of the circuit is left unchanged and its parameters are optimised to achieve smaller losses. The final loss of the optimised parameters is very close to 0.}
        \label{fig:vqls_4q_finetune}
    \end{subfigure}
    \caption{The search rewards and fine-tune loss for VQLS experiment. }\label{fig:vqls_search_finetune}
\end{figure}

\begin{figure}[H]
  \centering
  \begin{quantikz}[transparent, row sep={0.8cm,between origins}]
\qw & \gate{H} & \gate{Rot} & \targ{}\vqw{0} & \qw & \targ{}\vqw{0} & \gate{Rot} & \qw\\
\qw & \gate{H} & \gate{Rot} & \ctrl{-1} & \gate{Rot} & \ctrl{-1} & \gate{Rot} & \qw\\
\qw & \gate{H} & \gate{Rot} & \ctrl{0} & \qw & \qw & \qw & \qw\\
\qw & \gate{H} & \gate{Rot} & \targ{}\vqw{-1} & \qw & \qw & \qw & \qw
\end{quantikz}
  \caption{Circuit searched for the VQLS problem. $Rot(\phi, \theta, \omega)=R_Z(\omega) R_Y(\theta) R_Z(\phi)$. The four Hadamard gates at the beginning of the circuit are to put everything in an equal superposition, and not included when constructing the search tree, i.e. the composed circuits will always start with four Hadamard gates placed on the four qubits. When drawing the circuit, the Placeholder gates, which are just identity gates, are removed from searched $\mathcal{P}$, although they were considered when constructing the search tree.}
  \label{fig:vqls_circ}
\end{figure}
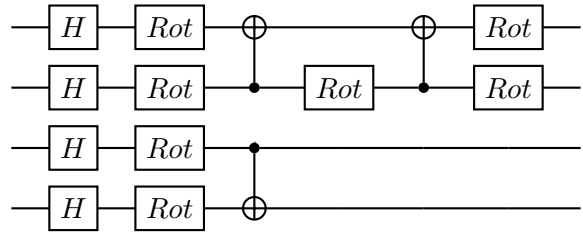

\begin{figure}[H]
  \centering
  \includegraphics[width=0.95\textwidth]{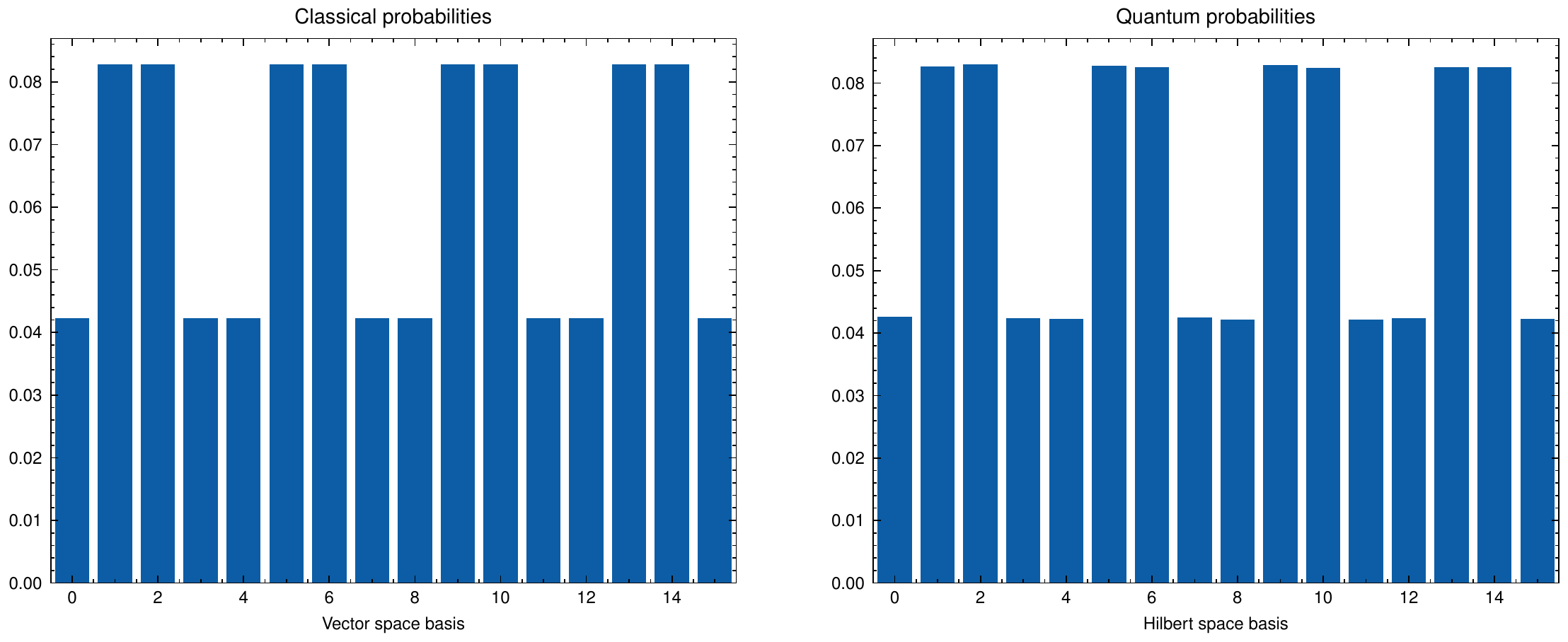}
  \caption{Comparison between classical probabilities, which obtained from solving the matrix equation with the classical method, i.e. $x = A^{-1}b$, of the normalised solution vector $\frac{x}{||x||}$ for $Ax = b$ (left), and the probabilities obtained by sampling the state $\ket{x}$ produced by the trained circuit in Fig~\ref{fig:vqls_circ} (right). The number of shots for measurement is $10^6$. We can see that the quantum results is very close to the classically obtained ones, showing that our algorithm can be indeed applied to finding variational ans\"atz for VQLS problems.}
  \label{fig:vqls_results_compare}
\end{figure}

The search rewards as well as fine-tune losses are shown in Fig~\ref{fig:vqls_search_finetune}. We can see that the search algorithm can produce a circuit with high reward (exceeds the threshold) quickly and the loss of the optimised parameters can reach close to 0. Although facing a large search space, our algorithm can still find a circuit (shown in Fig~\ref{fig:vqls_circ}) that minimises the loss function (Fig~\ref{fig:vqls_4q_finetune}) and leads us to results close to the classical solution. A comparison of the results obtained by directly solving the linear equation $Ax=b$ and the results obtained by sampling the state $\ket{x}$ produced by the searched circuit is shown in Fig~\ref{fig:vqls_results_compare}.

\subsection{Search for quantum chemistry ans\"atze}\label{h2}

Recently, there has been a lot of progress made on finding the ground state energy of a molecule on a quantum computer with variational circuit, both on theoretical \cite{li2017efficient,mcclean2016theory,wecker2015progress} and experimental \cite{peruzzo2014variational,o2016scalable, colless2017implementing, kandala2017hardware, colless2018computation, dumitrescu2018cloud} front. 
Normally, when designing the ans\"atz for the ground energy problem either a physically plausible or a hardware efficient ans\"atz needs to be found. However, our algorithm provides an approach which can minimise the effort needed to carefully choose an ans\"atz and automatically design the circuit according to the device gate set and topology.

Generally speaking, solving the ground energy problem with quantum computers is an application of the variational principle \cite{sakurai_napolitano_2017}:
\begin{equation}
E_0 \leq \frac{\langle\tilde{0}|H| \tilde{0}\rangle}{\langle\tilde{0} \mid \tilde{0}\rangle} \label{eq:variational}
\end{equation}
where $H$ is the system Hamiltonian, $| \tilde{0}\rangle$ is the ``trail ket'' \cite{sakurai_napolitano_2017}, or ans\"atz, trying to mimic the real wave function at ground state with energy $E_0$, which is the smallest eigenvalue of the system Hamiltonian H. Starting from $\ket{0^{\otimes n}}$ for an $n-$qubit system, the ``trial ket'' can be written as a function of a set of (real) parameters $\theta$:
\begin{equation}
    \ket{\tilde{0}} = \ket{\varphi (\theta)} = U(\theta)\ket{0^{\otimes n}}
\end{equation}
Given an ans\"atz, the goal of optimisation is to find a set of parameters $\theta$ that minimises the right hand side of Eqn \ref{eq:variational}. However, in our research, the form of the trail wave function will no longer be fixed. We will not only vary the parameters, but also the circuit structure that represent the ans\"atz.

\subsubsection{Experiment settings}

\paragraph{Search an ans\"atz for finding the ground energy of $\text{H}_2$:}
In this experiment, we adopted the 4-qubit Hamiltonian $H_{hydrogen}$ for the hydrogen molecule $\text{H}_2$ generated by the Pennylane-QChem \cite{bergholm2020pennylane} package, when the coordinates of the two hydrogen atoms are $(0, 0, -0.66140414)$ and $(0, 0, 0.66140414)$, respectively, in atom units. The goal of this experiment is to find an ans\"atz that can produce similar states as the four-qubit Givens rotation for single and double excitation.
The unitary operator \footnote{see \url{https://pennylane.readthedocs.io/en/latest/code/api/pennylane.SingleExcitation.html}} that performs single excitation on a subspace spanned by $\{\ket{01}, \ket{10} \}$ can be written as 
\begin{equation}
U(\phi)=\left[\begin{array}{cccc}
1 & 0 & 0 & 0 \\
0 & \cos (\phi / 2) & -\sin (\phi / 2) & 0 \\
0 & \sin (\phi / 2) & \cos (\phi / 2) & 0 \\
0 & 0 & 0 & 1
\end{array}\right]
\end{equation}
And the transformation of the double excitation on the subspace spanned by $\{|1100\rangle,|0011\rangle\}$ is\footnote{see \url{https://pennylane.readthedocs.io/en/latest/code/api/pennylane.DoubleExcitation.html}} :
\begin{equation}
\begin{aligned}
&|0011\rangle \rightarrow \cos (\phi / 2)|0011\rangle+\sin (\phi / 2)|1100\rangle \\
&|1100\rangle \rightarrow \cos (\phi / 2)|1100\rangle-\sin (\phi / 2)|0011\rangle
\end{aligned}
\end{equation}

Following \cite{pennylane_dev_team_2021}, we initialised the circuit with the 4-qubit vacuum state $\vert \psi_0\rangle=\vert 0000\rangle$. We denote the unitary for the searched ans\"atz $U_{\rm SearchedAnsatz}$, which is a function of its structure $\mathcal{P}_{\rm SearchedAnsatz}$ and corresponding parameters. Then the loss and reward functions can be written as:
\begin{equation}
    L_{\text{H}_2} =\langle \psi_0 \vert U_{\rm SearchedAnsatz}^{\dagger} H_{\rm Hydrogen} U_{\rm SearchedAnsatz} \vert \psi_0\rangle
\end{equation}
\begin{equation}
    R_{\text{H}_2} = -L_{\text{H}_2}
\end{equation}

The operation pool consists of Placeholder gates, Rot gates and CNOT gates with a linear entanglement topology (nearest neighbour interactions). The maximum number of layers is 30, with maximum number of CNOT gates $30/2 = 15$, and no penalty term for the number of Placeholder gates:
\begin{equation}
    \mathcal{R}_{\text{H}_2, \rm Pool\;1} = R_{\text{H}_2}
\end{equation}

Such settings of operation pool and number of layers will give us an overall search space of size $14^{30}\approx 2.42\times 10^{34}$. However, the imposed hard limits and gate limits will drastically reduce the size of the search space.

\paragraph{Search an ans\"atz for finding the ground energy of $\text{LiH}$}
The loss and reward functions for the $\text{LiH}$ task are similar to the $\text{H}_2$ one:
\begin{equation}
    L_{\text{LiH}} =\langle \psi_0 \vert U_{\rm SearchedAnsatz}^{\dagger} H_{\text{LiH}} U_{\rm SearchedAnsatz} \vert \psi_0\rangle
\end{equation}
\begin{equation}
    R_{\text{LiH}} = -L_{\text{LiH}}
\end{equation}
and the initial state is also the vacuum state $\ket{\psi_0} = \ket{0}^{\otimes 10}$. The Hamiltonian is obtained at bond length 2.969280527 Bohr, or 1.5712755873606 Angstrom, with 2 active electrons and 5 active orbitals. The size of the operation pool $c = \vert \mathcal{C} \vert = 38$, including Rot gates, Placeholder and CNOT gates operating on neighbouring qubits on a line topology. The maximum number of layers is 20, giving us a search space of size $\vert \mathcal{S} \vert = 38^{20} \approx 3.94\times 10^{31}$. The `hard limit' on the number of CNOT gates in the circuit is $20/2=10$.

\paragraph{Search an ans\"atz for finding the ground energy of $\text{H}_2\text{O}$} The loss and reward functions of the water molecule are shown as follows:
\begin{equation}
    L_{\text{H}_2\text{O}} =\langle \psi_0 \vert U_{\rm SearchedAnsatz}^{\dagger} H_{\text{H}_2\text{O}} U_{\rm SearchedAnsatz} \vert \psi_0\rangle
\end{equation}
\begin{equation}
    R_{\text{H}_2\text{O}} = -L_{\text{H}_2\text{O}}
\end{equation}
and the initial state is also the vacuum state $\ket{\psi_0} = \ket{0}^{\otimes 8}$. The Hamiltonian is obtained when the three atoms are positioned at the following coordinates:
\begin{equation}
    \text{H}:(0.,0.,0.); 
    \text{O}:(1.63234543, 0.86417176, 0);
    \text{H}:(3.36087791, 0.,0.)
\end{equation}
Units are in Angstrom.
Active electrons is set to 4 and active orbitals is set to 4. The size of the operation pool $c = \vert \mathcal{C} \vert = 30$, including Rot gates, Placeholder and CNOT gates operating on neighbouring qubits on a line topology. The maximum number of layers is 20, giving us a search space of size $\vert \mathcal{S} \vert = 30^{50} \approx 7.18\times 10^{73}$. The `hard limit' on the number of CNOT gates in the circuit is 25.
\subsubsection{Results}

\begin{figure}[H]
    \centering
    \begin{subfigure}[t]{0.48\textwidth}
        \includegraphics[width=\textwidth]{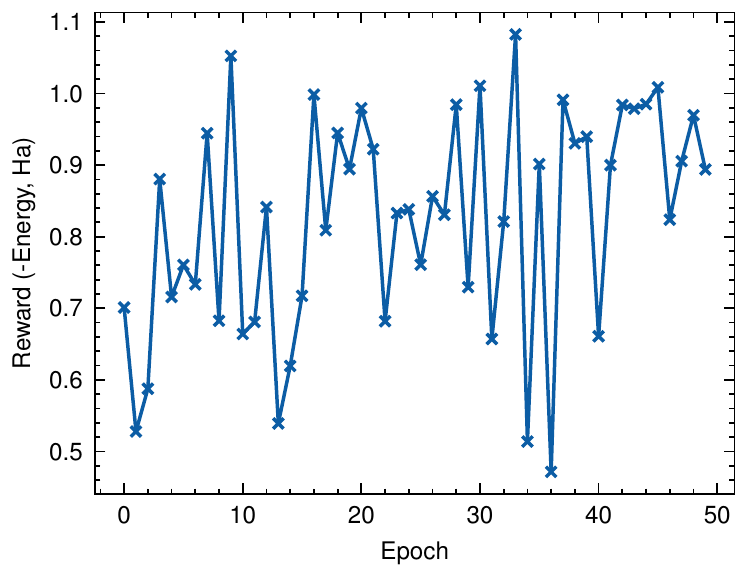}
        \caption{Search rewards for the $\text{H}_2$ ans\"atz. We can see that for most of the 50 iterations, the reward for the best circuit sampled from the search tree stays over 0.7.}
        \label{fig:h2_search}
    \end{subfigure}
    ~ 
    \begin{subfigure}[t]{0.48\textwidth}
        \includegraphics[width=\textwidth]{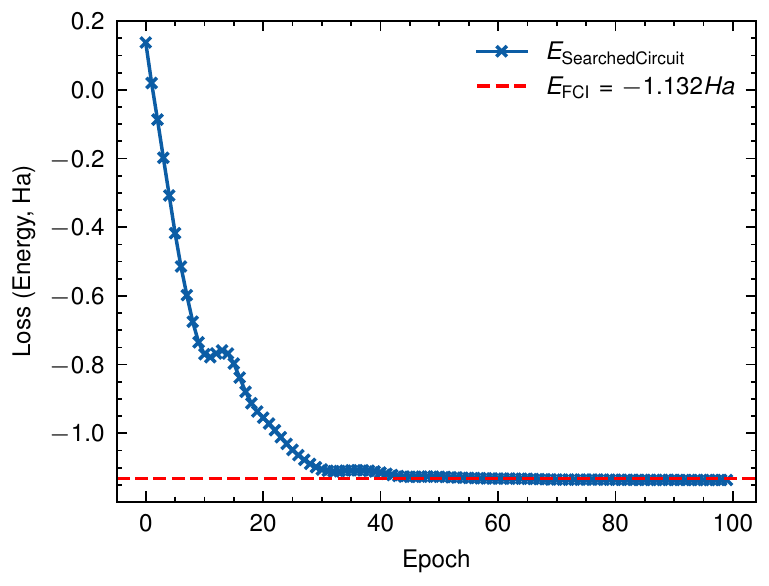}
        \caption{Fine-tune loss for the searched $\text{H}_2$ circuit. At the last iteration of optimisation, the energy is around -1.1359 Ha. The classically computed full configuration interaction result with PySCF \cite{Sun2018-nq, Sun2020-ej}, which is around -1.132 Ha and marked by the red horizontal dashed line. The difference between the energy achieved by the searched circuit and PySCF is close to chemical accuracy.)}
        \label{fig:h2_finetune}
    \end{subfigure}
    \caption{The search rewards and fine-tune loss for $\text{H}_2$ circuit. experiment.}\label{fig:h2_search_finetune}
\end{figure}

\begin{figure}[H]
  \centering
  \begin{quantikz}[transparent, row sep={0.8cm,between origins}]
\qw & \gate{Rot} & \qw & \qw & \qw & \ctrl{0} & \gate{Rot} & \targ{}\vqw{0} & \ctrl{0} & \targ{}\vqw{0} & \qw & \qw & \qw & \ctrl{0} & \qw & \qw\\
\qw & \qw & \qw & \qw & \targ{}\vqw{0} & \targ{}\vqw{-1} & \qw & \ctrl{-1} & \targ{}\vqw{-1} & \ctrl{-1} & \targ{}\vqw{0} & \gate{Rot} & \targ{}\vqw{0} & \targ{}\vqw{-1} & \qw & \qw\\
\qw & \gate{Rot} & \ctrl{0} & \gate{Rot} & \ctrl{-1} & \ctrl{0} & \gate{Rot} & \targ{}\vqw{0} & \gate{Rot} & \targ{}\vqw{0} & \ctrl{-1} & \qw & \ctrl{-1} & \targ{}\vqw{0} & \gate{Rot} & \qw\\
\qw & \qw & \targ{}\vqw{-1} & \gate{Rot} & \qw & \targ{}\vqw{-1} & \qw & \ctrl{-1} & \qw & \ctrl{-1} & \qw & \qw & \qw & \ctrl{-1} & \qw & \qw
\end{quantikz}
  \caption{The circuit for finding the ground energy of the $\text{H}_2$ molecule produced by the search algorithm. We can see that there are already familiar structures emerging, like the SWAP gate between the first two qubits in the middle and the Ising coupling gate-like structure right under the decomposed SWAP gate.}
  \label{fig:h2_circ}
\end{figure}
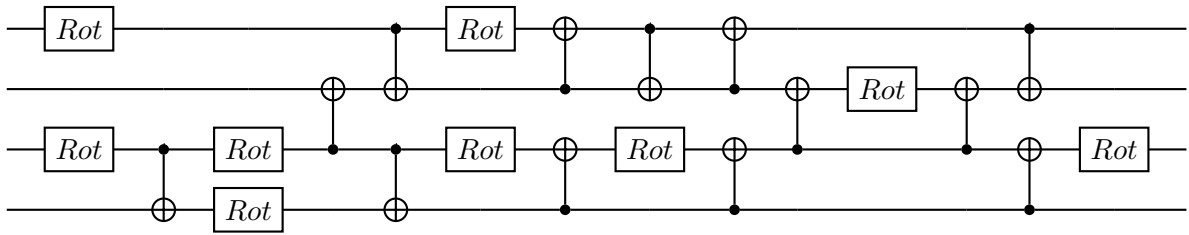

\paragraph{$\text{H}_2$ Results} The search reward when finding the suitable circuit structure is shown in Fig~\ref{fig:h2_search} and the training process for the circuit produced by the search algorithm is shown in Fig~\ref{fig:h2_finetune}. The ans\"atz is presented in Fig~\ref{fig:h2_circ}. We can see from Fig~\ref{fig:h2_circ} that the unitaries are not randomly placed on the four wires, instead there present familiar structures like the decomposition of the SWAP gate and Ising coupling gates. An example of the Ising coupling gates (often appears in quantum optimisation problems) is the $R_{ZZ}$ gate:
\begin{equation}
R_{Z Z}(\theta)=e^{ -i \frac{\theta}{2} Z \otimes Z}=\left[\begin{array}{cccc}
e^{-i \frac{\theta}{2}} & 0 & 0 & 0 \\
0 & e^{i \frac{\theta}{2}} & 0 & 0 \\
0 & 0 & e^{i \frac{\theta}{2}} & 0 \\
0 & 0 & 0 & e^{-i \frac{\theta}{2}}
\end{array}\right] = CNOT_{1,2}RZ_{2}(\theta)CNOT_{1,2}
\end{equation}
Where $CNOT_{1,2}$ is the CNOT gate controlled by the first qubit and target on the second qubit, and $RZ_{2}(\theta)$ is a Z-rotation gate on the second qubit.
However, other parts of the circuit are not familiar, which indicates that the search algorithm can go beyond human intuition. The total number of gates in the circuit is 22, including 13 local CNOT gates.

\begin{figure}[H]
    \centering
    \begin{subfigure}[t]{0.48\textwidth}
        \includegraphics[width=\textwidth]{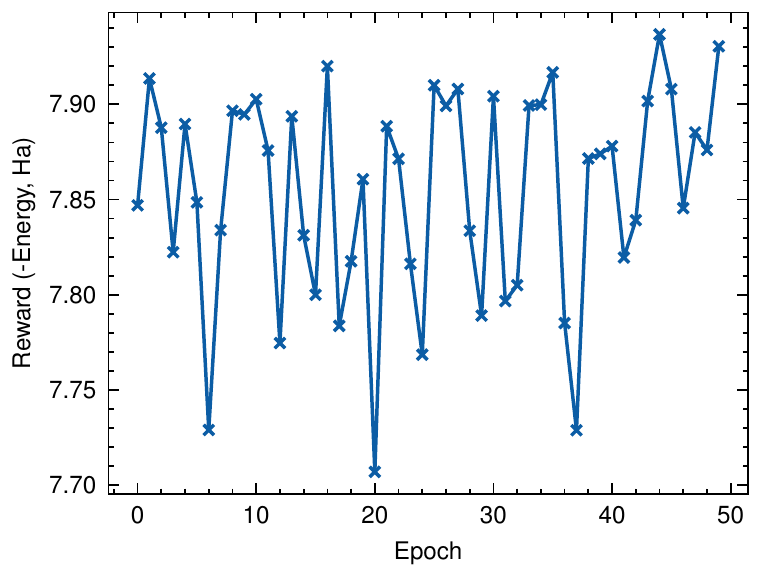}
        \caption{Search rewards for the $\text{LiH}$ ans\"atz. We can see that for most of the 50 iterations, the reward for the best circuit sampled from the search tree stays over 7.7.}
        \label{fig:lih_search}
    \end{subfigure}
    ~ 
    \begin{subfigure}[t]{0.48\textwidth}
        \includegraphics[width=\textwidth]{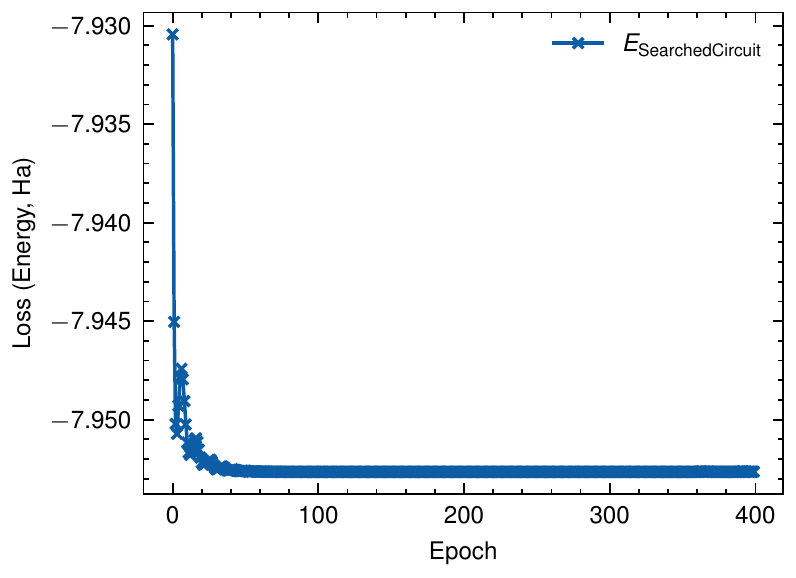}
        \caption{Fine-tune loss for the searched $\text{LiH}$ circuit. At the last iteration of optimisation, the energy is around -7.9526 Ha, close to the chemical accuracy compared to classically computed full configuration interaction energy with PySCF \cite{Sun2018-nq, Sun2020-ej}, which is around -7.8885 Ha}
        \label{fig:lih_finetune}
    \end{subfigure}
    \caption{The search rewards and fine-tune loss for $\text{LiH}$ circuit. experiment.}\label{fig:lih_search_finetune}
\end{figure}

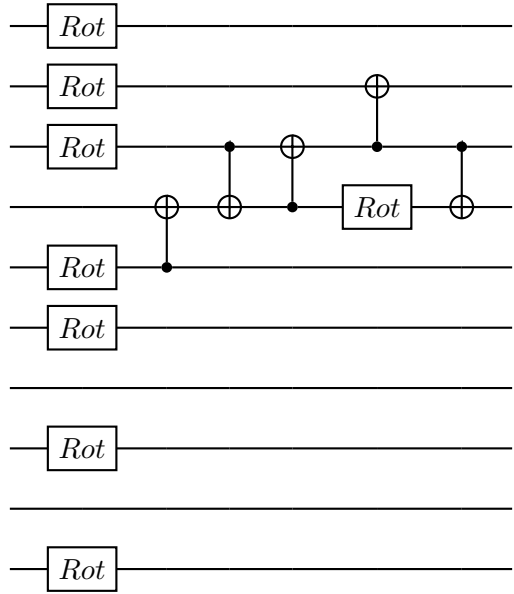
\begin{figure}[H]
  \centering
  \begin{quantikz}[transparent, row sep={0.8cm,between origins}]
\qw & \gate{Rot} & \qw & \qw & \qw & \qw & \qw & \qw\\
\qw & \gate{Rot} & \qw & \qw & \qw & \targ{}\vqw{0} & \qw & \qw\\
\qw & \gate{Rot} & \qw & \ctrl{0} & \targ{}\vqw{0} & \ctrl{-1} & \ctrl{0} & \qw\\
\qw & \qw & \targ{}\vqw{0} & \targ{}\vqw{-1} & \ctrl{-1} & \gate{Rot} & \targ{}\vqw{-1} & \qw\\
\qw & \gate{Rot} & \ctrl{-1} & \qw & \qw & \qw & \qw & \qw\\
\qw & \gate{Rot} & \qw & \qw & \qw & \qw & \qw & \qw\\
\qw & \qw & \qw & \qw & \qw & \qw & \qw & \qw\\
\qw & \gate{Rot} & \qw & \qw & \qw & \qw & \qw & \qw\\
\qw & \qw & \qw & \qw & \qw & \qw & \qw & \qw\\
\qw & \gate{Rot} & \qw & \qw & \qw & \qw & \qw & \qw
\end{quantikz}
  \caption{Circuit structure produced by the search algorithm for LiH. We can see that the structure of the circuit is quite simple, compared to the circuit for $\text{H}_2$ in Fig~\ref{fig:h2_circ}, indicating that the vacuum state $\ket{\psi_0} = \ket{0}^{\otimes 10}$ is already very close to the ground energy state.}
  \label{fig:lih_circ}
\end{figure}

\paragraph{$\text{LiH}$ Results} The search reward when finding the suitable circuit structure for LiH is shown in Fig~\ref{fig:lih_search} and the training process for the circuit produced by the search algorithm is shown in Fig~\ref{fig:lih_finetune}. The ans\"atz is presented in Fig~\ref{fig:lih_circ}. The circuit produced by the search algorithm is simpler compared to the $\text{H}_2$ ans\"atz in Fig~\ref{fig:h2_circ}, indicating that the initial state may be very close to the ground energy state.

\begin{figure}[H]
    \centering
    \begin{subfigure}[t]{0.48\textwidth}
        \includegraphics[width=\textwidth]{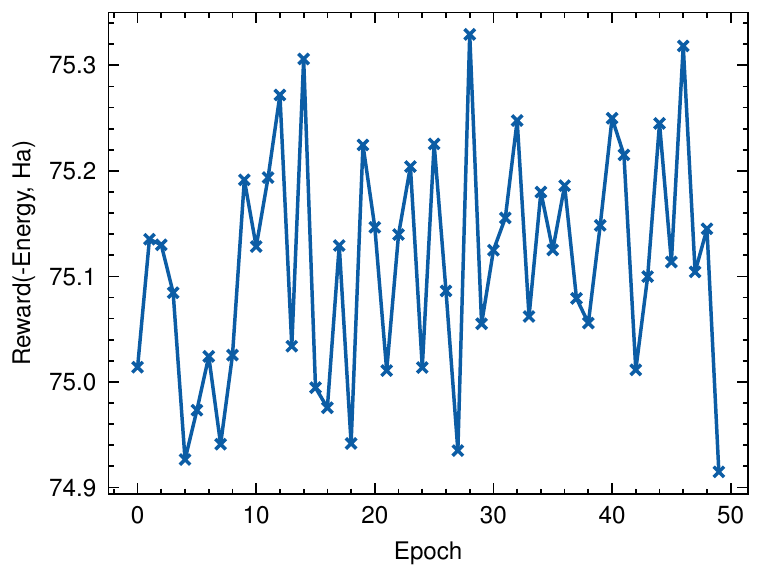}
        \caption{Search rewards for the $\text{H}_2\text{O}$ ans\"atz. We can see that for most of the 50 iterations, the reward for the best circuit sampled from the search tree stays over 74.9.}
        \label{fig:h2o_search}
    \end{subfigure}
    ~ 
    \begin{subfigure}[t]{0.48\textwidth}
        \includegraphics[width=\textwidth]{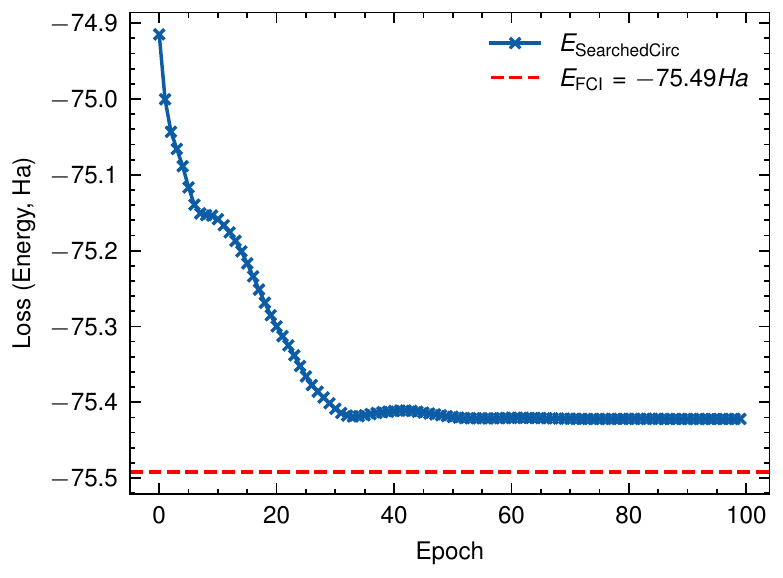}
        \caption{Fine-tune loss for the searched $\text{H}_2\text{O}$ circuit. At the last iteration of optimisation, the energy is around -75.4220  Ha,  close to the chemical accuracy compared to classically computed full configuration interaction energy with PySCF \cite{Sun2018-nq, Sun2020-ej}, which is around -75.4917  Ha}
        \label{fig:h2o_finetune}
    \end{subfigure}
    \caption{The search rewards and fine-tune loss for $\text{H}_2\text{O}$ circuit.}\label{fig:h2o_search_finetune}
\end{figure}

\begin{figure}[H]
  \centering
 \includegraphics[width=0.9\textwidth]{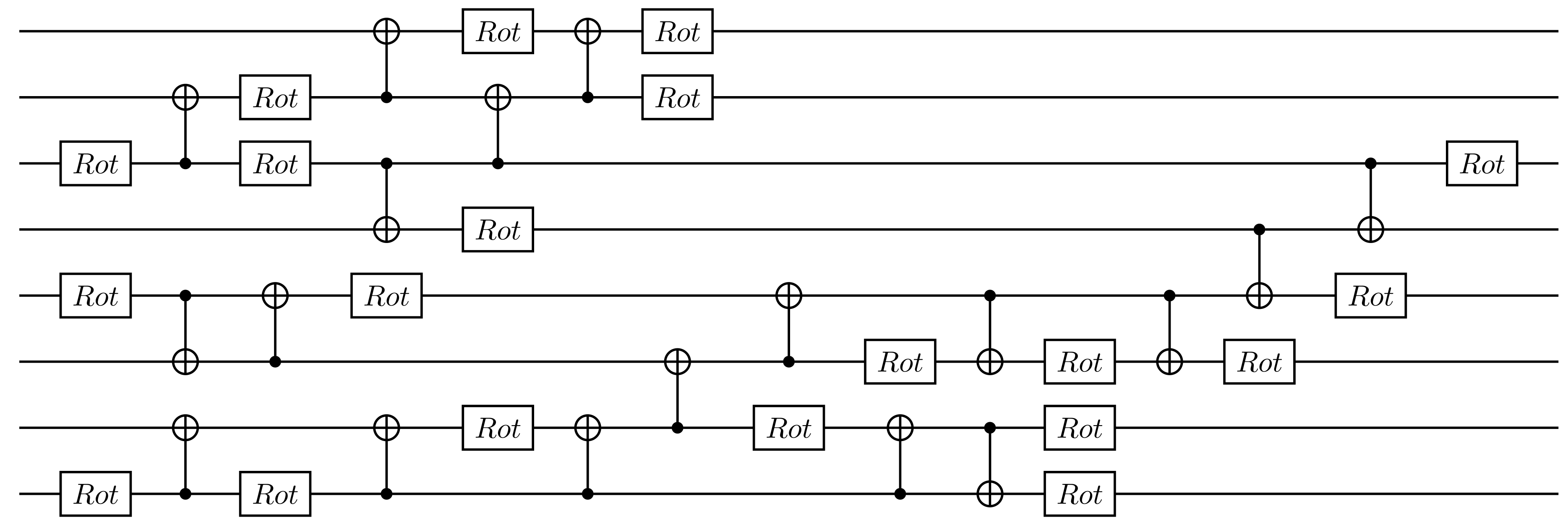}
  \caption{Circuit for $\text{H}_2\text{O}$ produced by the search algorithm.}
  \label{fig:h2o_circ}
\end{figure}

\paragraph{$\text{H}_2\text{O}$ Results} The search reward when finding the suitable circuit structure for $\text{H}_2\text{O}$ is shown in Fig~\ref{fig:h2_search} and the training process for the circuit produced by the search algorithm is shown in Fig~\ref{fig:h2o_finetune}. The ans\"atz is presented in Fig~\ref{fig:h2o_circ}, which has 38 gates in total, including 10 local CNOT gates. Although there are still some familiar structures, such as the Ising coupling in the circuit, the heuristics behind most parts of the circuit are already unintuitive for human researchers.

\subsection{Solving the \textsc{MaxCut} problem}
As a classic and well-known optimisation problem, the \textsc{MaxCut} problem plays an important role in network science, circuit design, as well as physics \cite{Bharti2022-sw}. The objective of the \textsc{MaxCut} problem is to find a partition $z$ of vertices in a graph $G = (V, E)$ which maximises the number of edges connecting the vertices in two disjoint sets $A$ and $B$:
\begin{equation}
    C(z) =\sum_{a=1}^m C_a(z)
\end{equation}
where $C_a(z) = 1$ if the $a^{th}$ edge connects one vortex in set $A$ and one vortex in set $B$, and $C_a(z) = 0$ otherwise. To perform the optimisation on a quantum computer, we will need to transform the cost function into Ising formulation:
\begin{equation}
    H_C = -\sum_{(i, j)\in E} \frac{1}{2} (I - Z_i Z_j)w_{ij}
\end{equation}\label{qaoa_ham}
where $Z_i$ is the Pauli $Z$ operator on the $i^{th}$ qubit and $w_{ij}$ is the weight of edge $(i, j)\in E$ for weighted \textsc{MaxCut} problem. For unweighted problems, $w_{ij} = 1$. In this formulation, vertices are represented by qubits in computational bases. By finding the wave-function that minimises the cost Hamiltonian $H_C$, we can find the solution that maximises $C(z)$. Previously, the major components of the QAOA (quantum approximate optimisation algorithm) ans\"atz are the cost Hamiltonian encoded by the cost unitary and the mixing Hamiltonians encoded by the mixing unitaries \cite{Farhi2014-ug}. Although this ans\"atz can find all the solutions in a equal superposition form, it is not always effective when the number of layers is small. Also, when the number of qubits (vertices) grows, the required number of layers and the number of shots during measurement to extract all of the solutions will also grow.

Since we have already had a Hamiltonian as our cost function in Sec.~\ref{h2}, we follow similar approach as quantum chemistry to find one of the solutions when the number of vertices is large.

\subsubsection{Experiment Settings}
\paragraph{Unweighted \textsc{MaxCut}}
\begin{figure}[H]
  \centering
  \includegraphics[width=0.8\textwidth]{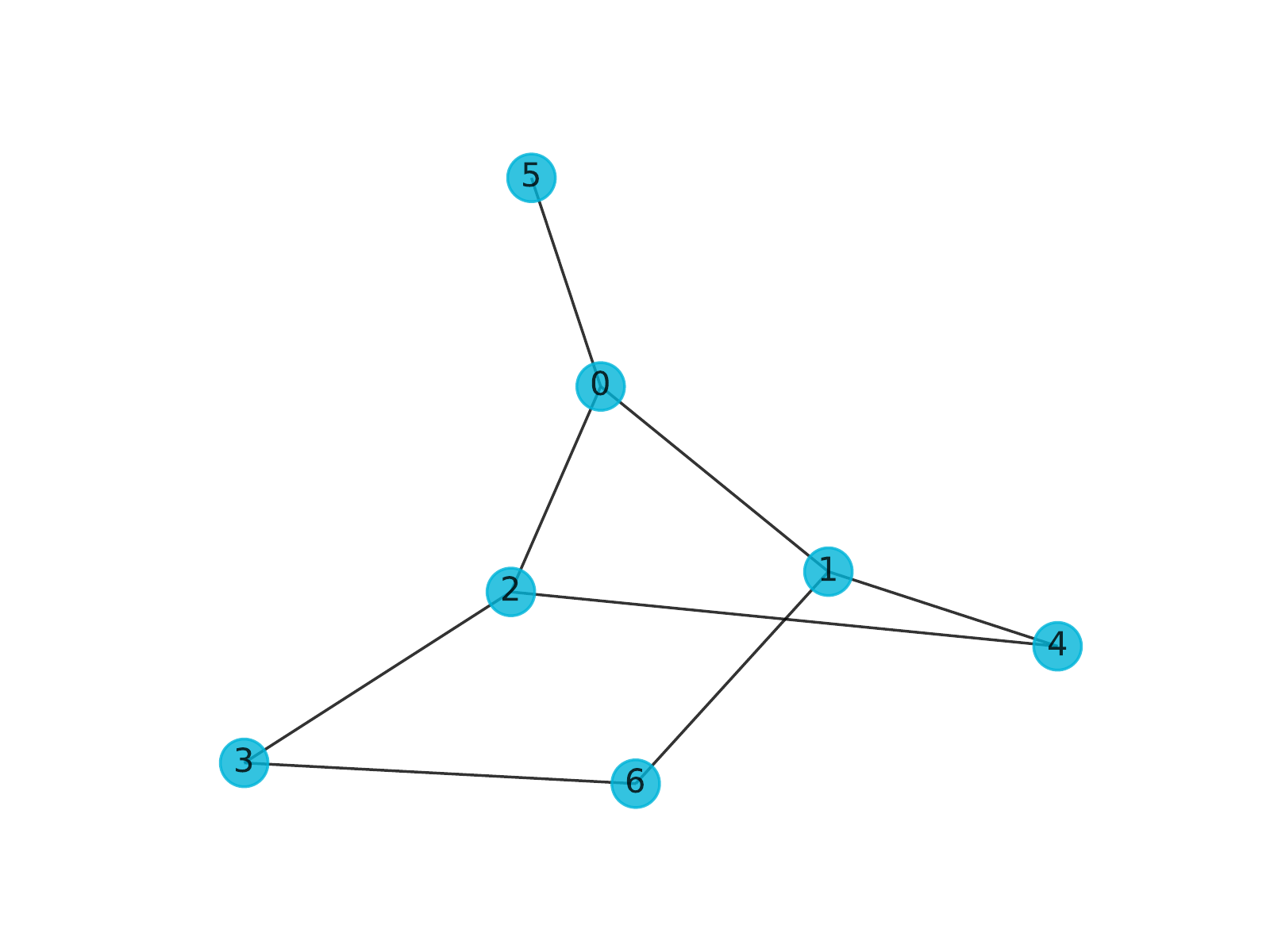}
  \caption{Problem graph for the unweighted \textsc{MaxCut} experiment}
  \label{fig:max_cut_prob}
\end{figure}
The problem graph for the unweighted \textsc{MaxCut} experiment is shown in Fig. \ref{fig:max_cut_prob}. This problem has six equally optimal solutions: 1001100, 0110010, 0111010, 1000101, 1001101 and 0110011, all have $C(z)=7$. The loss function is based on the expectation of the cost Hamiltonian $H_C$:
\begin{equation}
    L_{\textsc{MaxCut}} = (\bra{+})^{\otimes 7}U_{\rm SearchedAnsatz}^{\dagger} H_C U_{\rm SearchedAnsatz}   (\ket{+})^{\otimes 7}
\end{equation}
The reward function is simply the negative of the loss function:
\begin{equation}
    \mathcal{R}_{\textsc{MaxCut}} = -L_{\textsc{MaxCut}}
\end{equation}
We ran the search algorithm twice with the same basic settings, including the operation pool and the maximum number of layers. Since there is a random sampling process during the warm-up stage, the final solutions found by the algorithm are expected to be different. The operation pool consists of CNOT gates between every two qubits, the Placeholder and the single qubit rotation gate \cite{nielsen00}:
\begin{equation}
Rot(\phi, \theta, \omega)=R_Z(\omega) R_Y(\theta) R_Z(\phi)=\left[\begin{array}{cc}
e^{-i(\phi+\omega) / 2} \cos (\theta / 2) & -e^{i(\phi-\omega) / 2} \sin (\theta / 2) \\
e^{-i(\phi-\omega) / 2} \sin (\theta / 2) & e^{i(\phi+\omega) / 2} \cos (\theta / 2)
\end{array}\right]
\end{equation}
The size of the operation pool $c = \vert \mathcal{C} \vert = 28$, and the number of layers $p = 15$, leading to a search space of size $\vert \mathcal{S} \vert = 28^{15} \approx 5 \times10^{21}$. The `hard' restrictions on the maximum number of CNOT gates in a circuit, which is 7,  can help reduce the size of the search space.

\begin{figure}[H]
  \centering
  \includegraphics[width=0.8\textwidth]{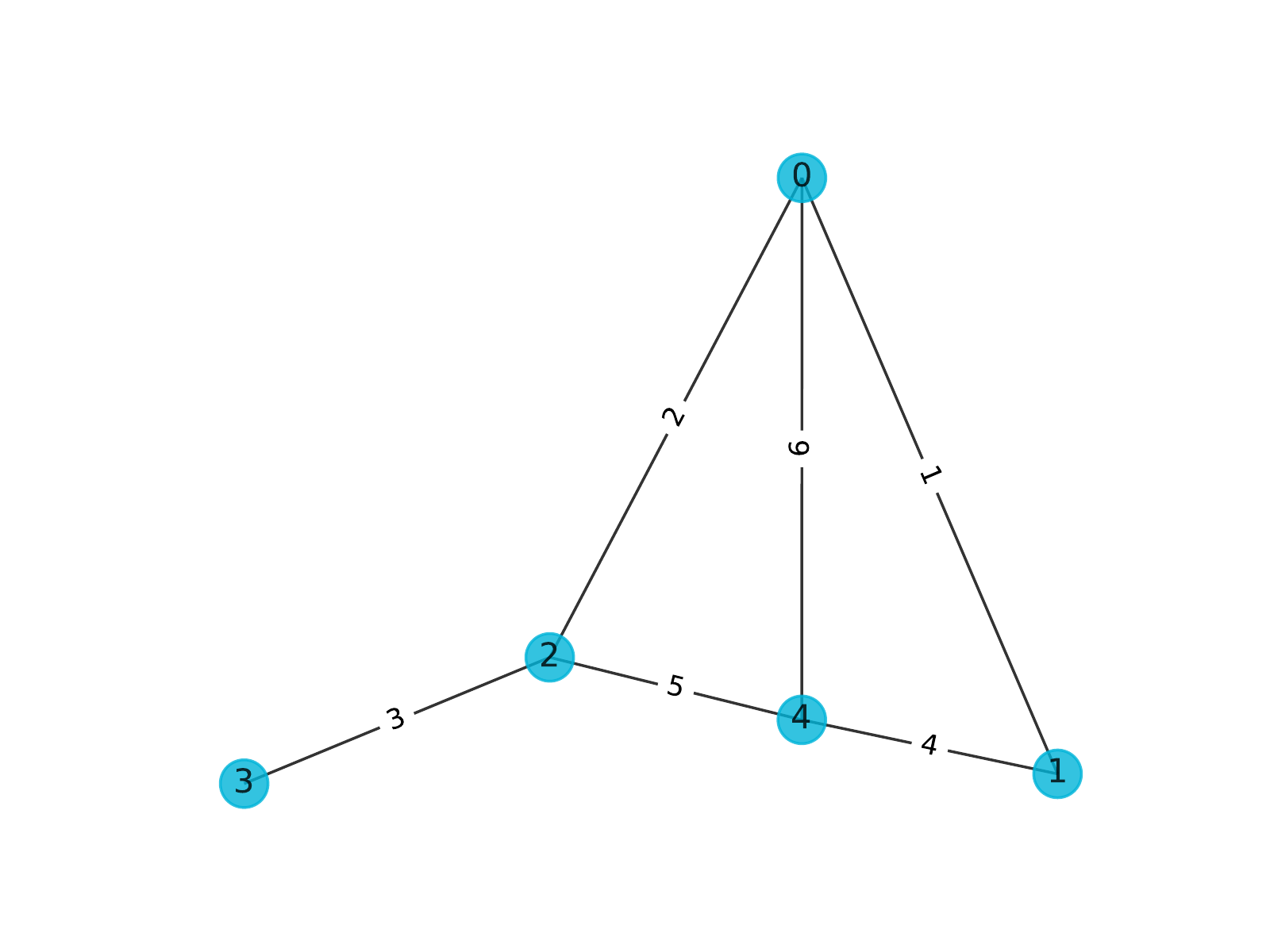}
  \caption{Problem graph for the weighted \textsc{MaxCut} experiment. The weights on edges (0,2), (0,4), (0,1), (2,4), (4,1) and (2,3) are 2, 6, 1, 5, 4 and 3, respectively.}
  \label{fig:max_cut_weighted_prob}
\end{figure}

\paragraph{Weighted \textsc{MaxCut}} For weighted \textsc{MaxCut}, we have a five-node graph, which is shown in Fig~\ref{fig:max_cut_weighted_prob}. The solution for this problem, 00011 (11100) is simpler than the unweighted version. The reward and loss function follow the same principle of the unweighted problem. The size of the operation pool $c = \mathcal{C} = 20$, and the number of layers $p = 10$, leading to a search space of size $\vert \mathcal{S} \vert = 20^{10} \approx 1.02\times 10^{13}$. The `hard' restriction on the maximum number of CNOTs in the circuit is 5.

\subsubsection{Results}
\begin{figure}[H]
    \centering
    \begin{subfigure}[b]{0.48\textwidth}
    \centering
        \begin{quantikz}[transparent, row sep={0.8cm,between origins}]
\qw & \gate{H} & \gate{Rot} & \qw & \qw & \qw & \qw & \qw\\
\qw & \gate{H} & \gate{Rot} & \qw & \qw & \qw & \qw & \qw\\
\qw & \gate{H} & \gate{Rot} & \ctrl{0} & \gate{Rot} & \targ{}\vqw{0} & \ctrl{0} & \qw\\
\qw & \gate{H} & \qw & \targ{}\vqw{-1} & \gate{Rot} & \ctrl{-1} & \targ{}\vqw{-1} & \qw\\
\qw & \gate{H} & \gate{Rot} & \qw & \qw & \qw & \qw & \qw\\
\qw & \gate{H} & \gate{Rot} & \qw & \qw & \qw & \qw & \qw\\
\qw & \gate{H} & \gate{Rot} & \qw & \qw & \qw & \qw & \qw
\end{quantikz}
        \caption{}
        \label{fig:qaoa_7q_first_circ}
    \end{subfigure}
    ~ 
    \begin{subfigure}[b]{0.48\textwidth}
    \centering
       \begin{quantikz}[transparent, row sep={0.8cm,between origins}, column sep = 0.2cm]
\qw & \gate{H} & \qw & \targ{}\vqw{0} & \gate{Rot} & \qw & \qw & \qw & \qw & \qw & \qw\\
\qw & \gate{H} & \gate{Rot} & \qw & \qw & \qw & \qw & \qw & \qw & \qw & \qw\\
\qw & \gate{H} & \qw & \qw & \gate{Rot} & \qw & \qw & \qw & \qw & \qw & \qw\\
\qw & \gate{H} & \qw & \qw & \gate{Rot} & \qw & \qw & \qw & \qw & \qw & \qw\\
\qw & \gate{H} & \qw & \qw & \gate{Rot} & \ctrl{0} & \targ{}\vqw{0} & \qw & \qw & \qw & \qw\\
\qw & \gate{H} & \qw & \qw & \gate{Rot} & \targ{}\vqw{-1} & \ctrl{-1} & \targ{}\vqw{0} & \ctrl{0} & \gate{Rot} & \qw\\
\qw & \gate{H} & \gate{Rot} & \ctrl{-6} & \qw & \qw & \qw & \ctrl{-1} & \targ{}\vqw{-1} & \qw & \qw
\end{quantikz}
        \caption{}
        \label{fig:qaoa_7q_second_circ}
    \end{subfigure}
    \caption{Two different circuits finding two different solutions of the \textsc{MaxCut} problem shown in Fig. \ref{fig:max_cut_prob}. Fig. \ref{fig:qaoa_7q_first_circ} gives the solution 0110010 (see Fig. \ref{fig:qaoa_7q_first_solution}) and Fig. \ref{fig:qaoa_7q_second_circ} gives the solution 0111010 (see Fig. \ref{fig:qaoa_7q_second_solution}).}\label{fig:qaoa_7q_circ}
\end{figure}
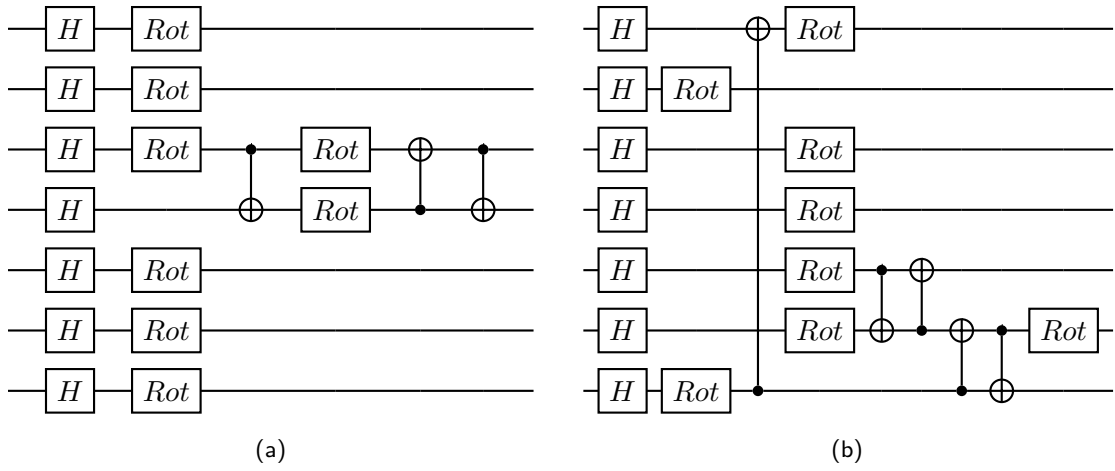

\begin{figure}[H]
    \centering
    \begin{subfigure}[b]{0.48\textwidth}
        \includegraphics[width=\textwidth]{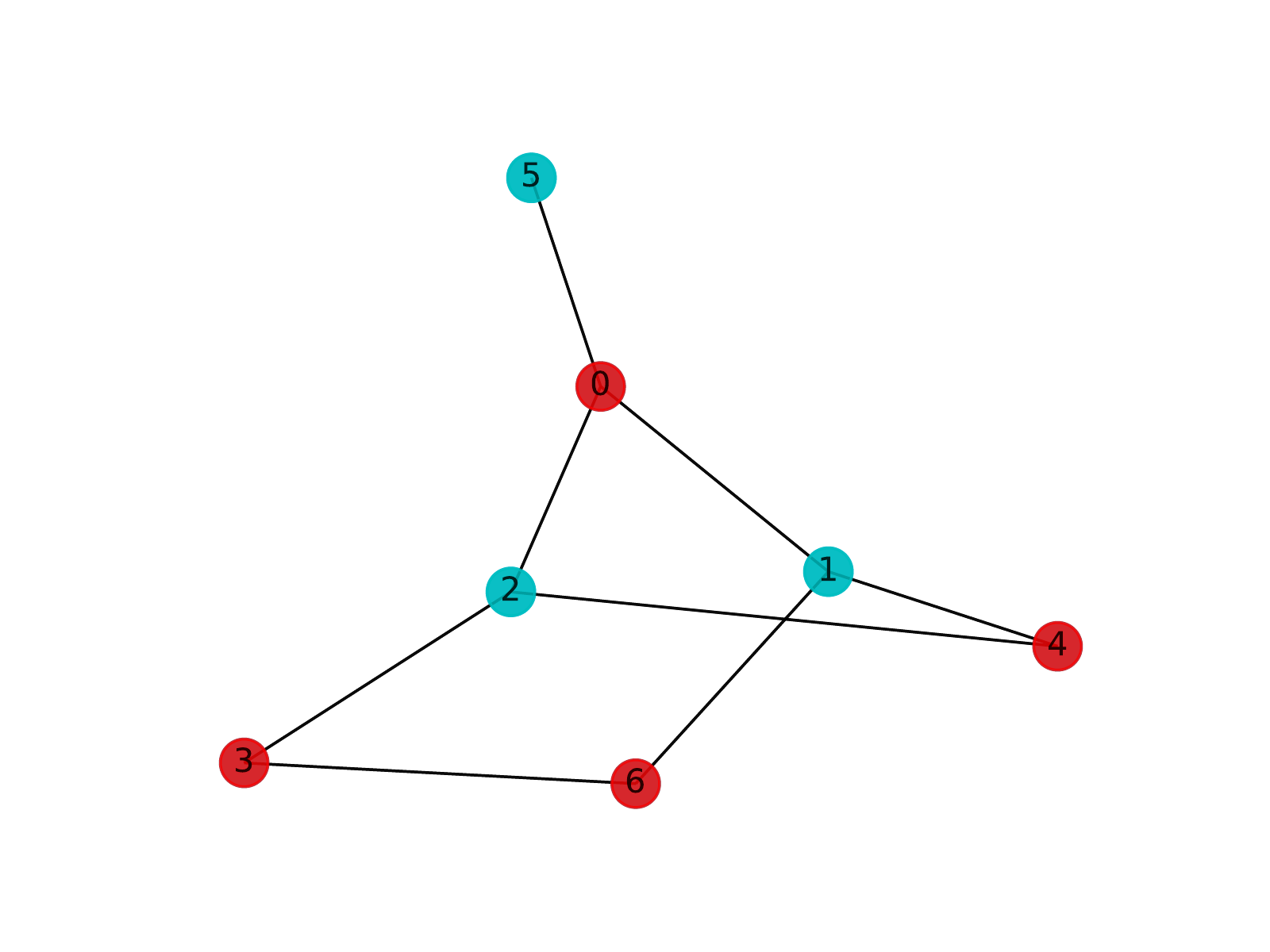}
        \caption{}
        \label{fig:qaoa_7q_first_solution}
    \end{subfigure}
    ~ 
    \begin{subfigure}[b]{0.48\textwidth}
        \includegraphics[width=\textwidth]{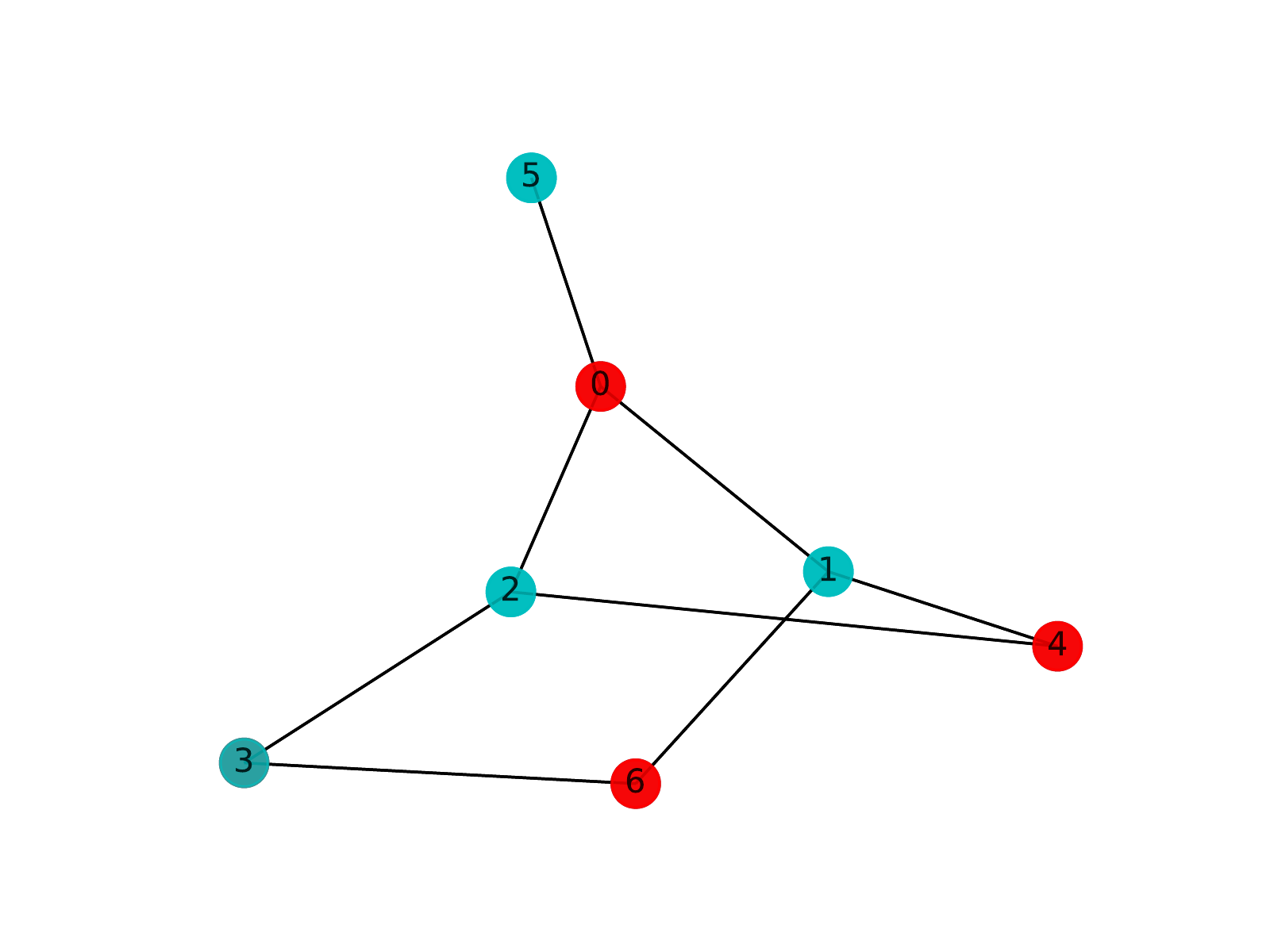}
        \caption{}
        \label{fig:qaoa_7q_second_solution}
    \end{subfigure}
    \caption{Two different optimal solutions found by the circuits in Fig. \ref{fig:qaoa_7q_first_circ} and Fig. \ref{fig:qaoa_7q_second_circ}, respectively.}\label{fig:qaoa_7q_solution}
\end{figure}

\begin{figure}[H]
    \centering
    \begin{subfigure}[t]{0.48\textwidth}
        \includegraphics[width=0.95\textwidth]{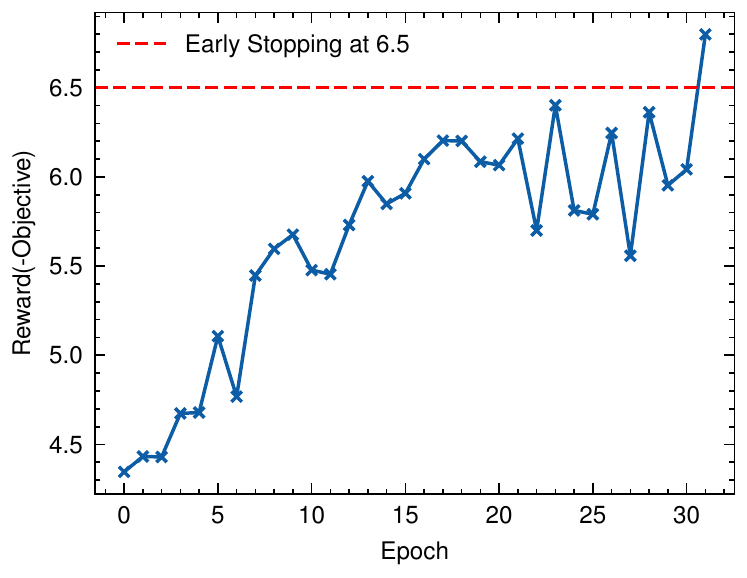}
        \caption{The change of rewards w.r.t. search iteration during the search for the ans\"atz (in Fig.\ref{fig:qaoa_7q_first_circ}) that gives the solution 0110010 (Fig. \ref{fig:qaoa_7q_first_solution}). To reduce the amount of time for searching, we stopped the algorithm after the search reward exceeded 6.5.}
        \label{fig:qaoa_search_reward_1}
    \end{subfigure}
    ~
    \begin{subfigure}[t]{0.48\textwidth}
        \includegraphics[width=\textwidth]{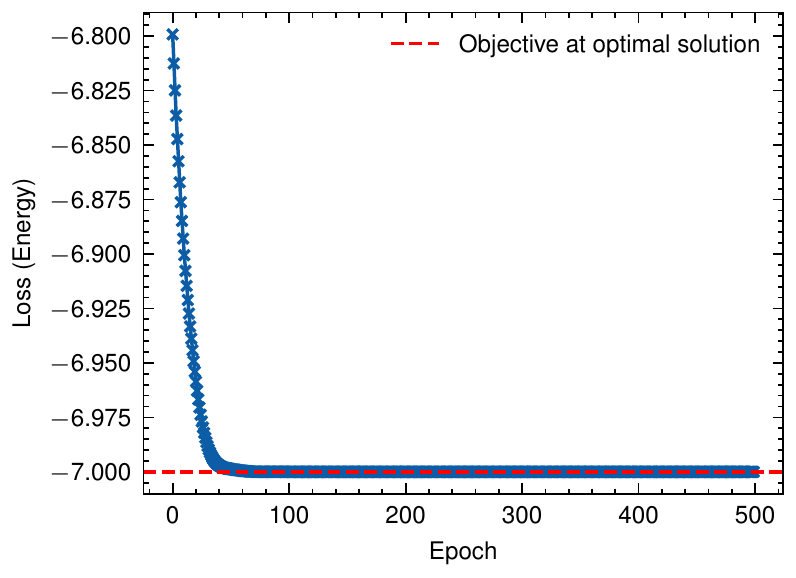}
        \caption{The change of loss w.r.t. optimisation iteration during the fine-tune for the ans\"atz (in Fig.\ref{fig:qaoa_7q_first_circ}) that gives the solution 0110010 (Fig. \ref{fig:qaoa_7q_first_solution}). We can see that the final loss is very close to -7, indicating that the circuit we found can produce an optimal solution.}
        \label{fig:qaoa_finetune_1}
    \end{subfigure}
    \hfill
    \begin{subfigure}[b]{0.48\textwidth}
        \includegraphics[width=0.99\textwidth]{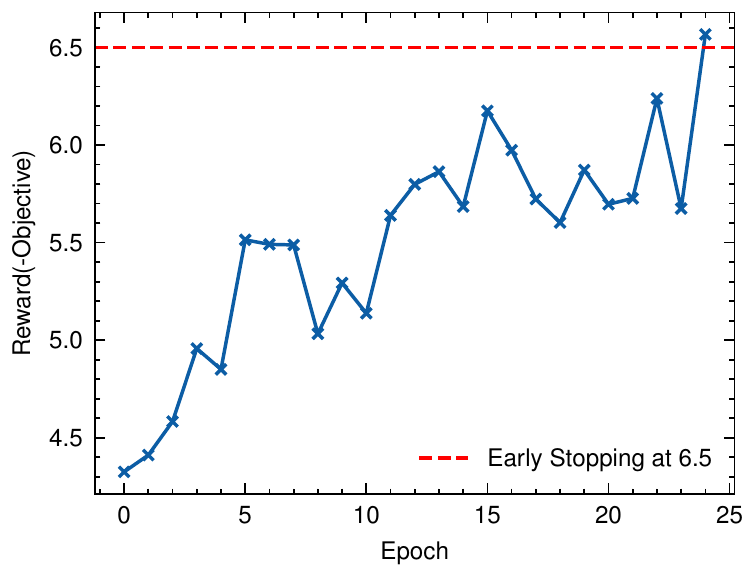}
        \caption{The change of rewards w.r.t. search iteration during the search for the ans\"atz (in Fig.\ref{fig:qaoa_7q_second_circ}) that gives the solution 0111010 (Fig. \ref{fig:qaoa_7q_second_solution}). To reduce the amount of time for searching, we stopped the algorithm after the search reward exceeded 6.5.}
        \label{fig:qaoa_search_reward_2}
    \end{subfigure}
    ~
    \begin{subfigure}[b]{0.48\textwidth}
        \includegraphics[width=\textwidth]{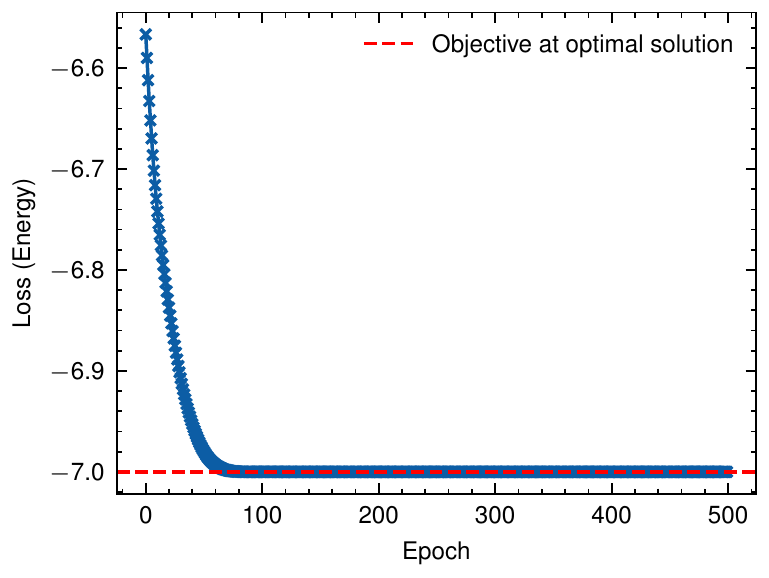}
        \caption{The change of loss w.r.t. optimisation iteration during the fine-tune for the ans\"atz (in Fig.\ref{fig:qaoa_7q_second_circ}) that gives the solution 0111010 (Fig. \ref{fig:qaoa_7q_second_solution}). We can see that the final loss is very close to -7, indicating that the circuit we found can produce an optimal solution.}
        \label{fig:qaoa_finetune_2}
    \end{subfigure}
    \caption{Search and fine-tune rewards for the circuits in Fig~\ref{fig:qaoa_7q_circ}. }\label{fig:qaoa_7q_search_finetune_both}
\end{figure}

\paragraph{Unweighted \textsc{MaxCut}}
The two runs of the search algorithms gave us two circuits (Fig. \ref{fig:qaoa_7q_circ}), leading to two of the six optimal solutions (Fig. \ref{fig:qaoa_7q_solution}). The search rewards and fine-tune losses for both circuits are shown in Figure~\ref{fig:qaoa_7q_search_finetune_both}. During the search stage, since we already know the maximum reward it could reach is 7, and the reward can only be integers, we set the early-stopping limit to 6.5 to reduce the amount of time spent on searching, which means the algorithm will stop searching and proceed to fine-tuning the parameters in the circuit after the reward exceeds 6.5. In a real-world application, we could let the search algorithm run through all of the pre-set number of iterations and record the best circuit structure as well as the corresponding rewards at each iteration at the same time. Then after the search stage finishes, we can choose the best circuit (or top-k circuits) in the search history to fine-tune, increasing our chance to find the optimal solution.

\begin{figure}[H]
    \centering
    \begin{subfigure}[b]{0.46\textwidth}
        \includegraphics[width=\textwidth]{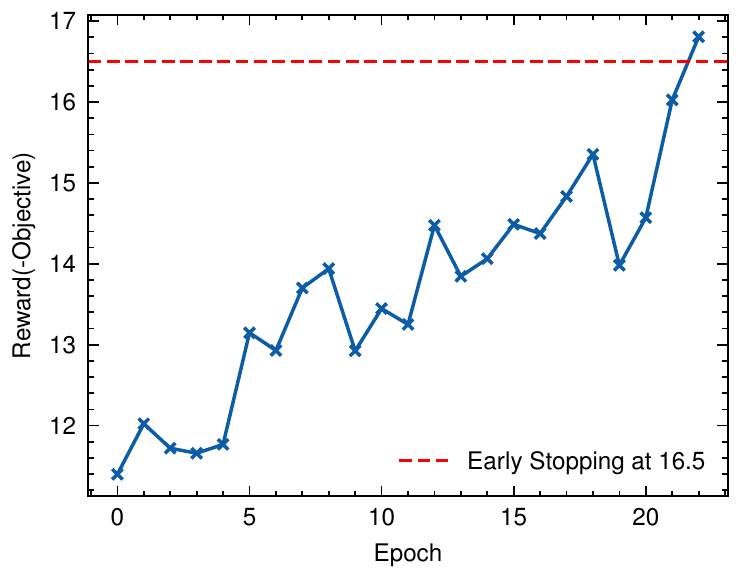}
        \caption{Search rewards for the five-node weighted \textsc{MaxCut} problem}
        \label{fig:qaoa_5q_search}
    \end{subfigure}
    ~ 
    \begin{subfigure}[b]{0.48\textwidth}
        \includegraphics[width=\textwidth]{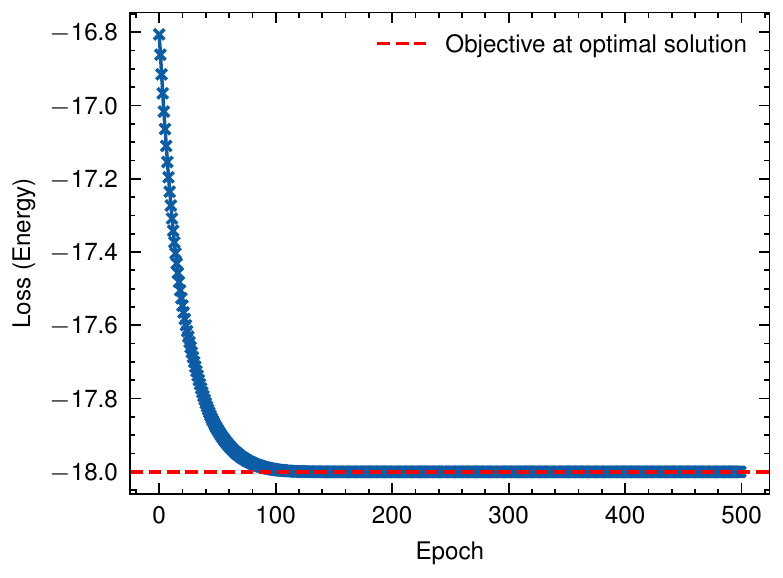}
        \caption{Fine-tune loss for the five-node weighted \textsc{MaxCut} problem}
        \label{fig:qaoa_5q_finetune}
    \end{subfigure}
    \caption{The search rewards and fine-tune losses of for the five-node \textsc{MaxCut} problem.}\label{fig:qaoa_5q_search_and_finetune}
\end{figure}

\begin{figure}[H]
    \centering
    \begin{subfigure}[b]{0.48\textwidth}
        \begin{quantikz}[transparent, row sep={0.8cm,between origins}]
\qw & \gate{H} & \qw & \targ{}\vqw{0} & \gate{Rot} & \targ{}\vqw{0} & \qw & \qw & \qw & \qw\\
\qw & \gate{H} & \qw & \qw & \gate{Rot} & \qw & \ctrl{0} & \qw & \qw & \qw\\
\qw & \gate{H} & \qw & \qw & \qw & \qw & \targ{}\vqw{-1} & \gate{Rot} & \ctrl{0} & \qw\\
\qw & \gate{H} & \qw & \qw & \qw & \qw & \gate{Rot} & \qw & \targ{}\vqw{-1} & \qw\\
\qw & \gate{H} & \gate{Rot} & \ctrl{-4} & \gate{Rot} & \ctrl{-4} & \qw & \qw & \qw & \qw
\end{quantikz}
        \caption{The searched circuit for the five-node weighted \textsc{MaxCut} problem}
        \label{fig:qaoa_5q_circ}
    \end{subfigure}
    ~ 
    \begin{subfigure}[b]{0.46\textwidth}
        \includegraphics[width=\textwidth]{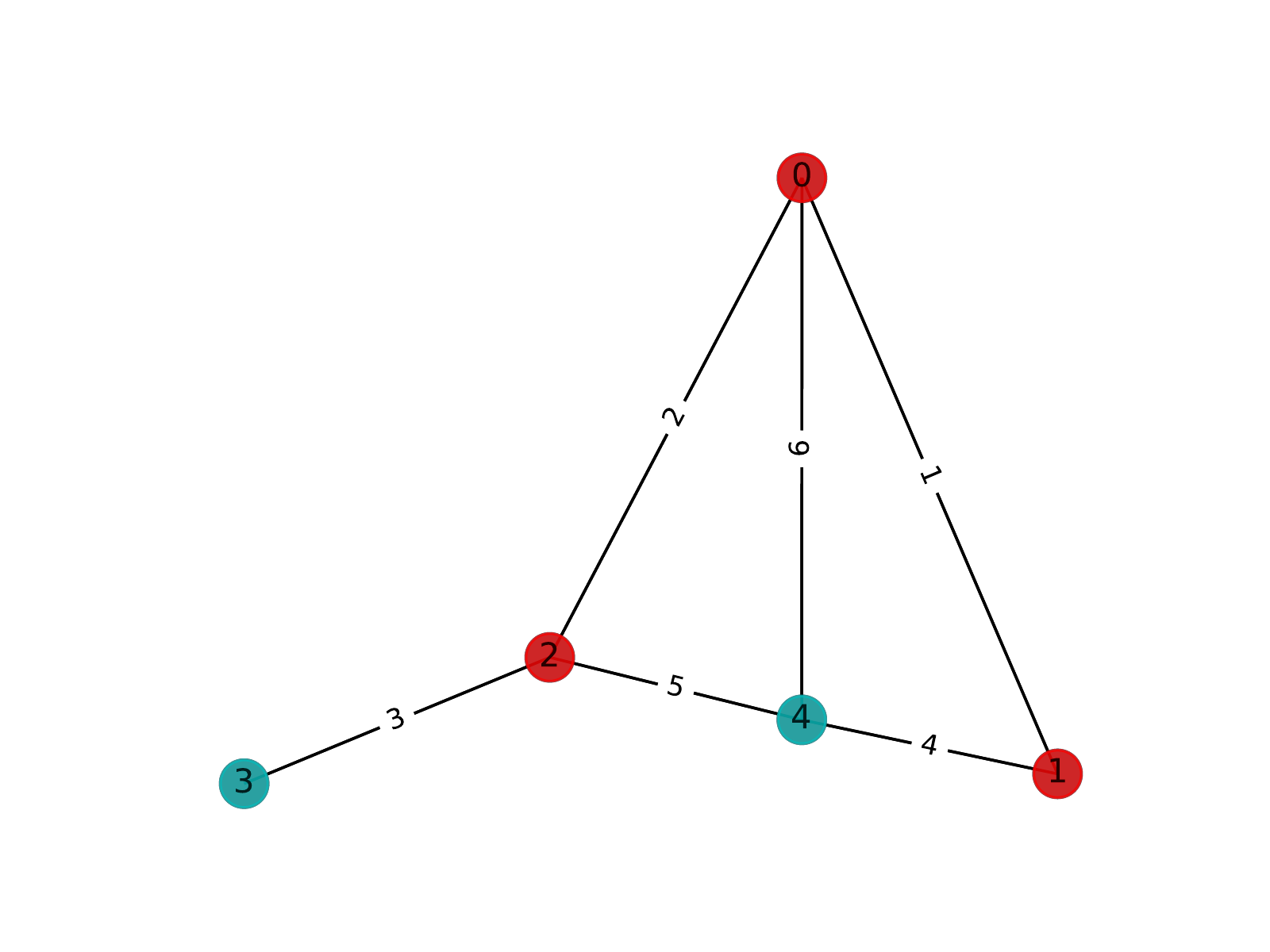}
        \caption{The solution sampled, which is 00011, from the circuit shown left.}
        \label{fig:qaoa_5q_solution}
    \end{subfigure}
    \caption{The searched circuit and sampled solution for the five-node \textsc{MaxCut} problem.}\label{fig:qaoa_5q_circ_and_solution}
\end{figure}

\paragraph{Weighted \textsc{MaxCut}}
The search rewards and fine-tune losses for the weighted \textsc{MaxCut} problem are shown in Fig~\ref{fig:qaoa_5q_search_and_finetune}. We can see that the search converged quickly and the fine-tune loss is very close to -18, indicating that the circuit (see Fig~\ref{fig:qaoa_5q_circ}) produced by our search algorithm can indeed find an optimal solution (see Fig~\ref{fig:qaoa_5q_solution}).

\section{Discussion}\label{discussion}
In this paper, we first formulated the circuit search problem as the tree structure. The sampled circuit can be represented as an arc (path from the root to a leaf) on the tree. We also introduced combinatorial multi-armed bandit and na\"ive assumption to model the selection of unitary operators for each layer in the circuit, and approximate the rewards of different unitaries with the reward of a fully constructed circuit. The search process is solved with Monte Carlo tree search (MCTS) algorithm. We demonstrated the effectiveness of our algorithmic framework with various examples, including finding the encoding circuit of the [[4,2,2]] quantum error detection code, developing the ans\"atz for variationally solving system of linear equations, searching the circuit for solving the ground state energy problem of different molecules, as well as circuits for solving optimisation problems on a graph. To our understanding, we are the first to propose such a versatile framework for the automated discovery of quantum circuits with MCTS and combinatorial multi-armed bandits. Results showed that our framework can be applied to many different areas, especially those with problems that can be formulated as finding the ground state energy of a certain Hamiltonian.

From the experiments and results shown in the previous sections, we can see that, by formulating quantum ans\"atz search as a tree-based search problem, one can easily impose various kinds of restrictions (`hard limits') on the circuit structure, leading to the pruning of the search tree and the search space. Also, by introducing Placeholders, one can explore smaller circuit sizes. Since current deep reinforcement learning algorithms struggle when the state space is large but the number of reward states is small. Compared to other research work in quantum ans\"atz search, including the differentiable quantum ans\"atz algorithm proposed in \cite{zhang2021differentiable}, and other QAS algorithms based on meta-learning \cite{chen2021quantum} or reinforcement learning \cite{kuo2021quantum}, which only investigate small-scale problems, like 3- or 4-qubit quantum Fourier transform in \cite{zhang2021differentiable}, 3-qubit classification task and 4-qubit $\text{H}_2$ ground state energy problem in \cite{du2020quantum}. A larger example can be seen in \cite{zhang2021neural}, which is a 6-qubit transversal Ising field model. In our research, we not only looked into 4-qubit systems like the $\text{H}_2$ molecule, but also larger systems like the $\text{LiH}$ and $\text{H}_2\text{O}$ molecule as well as \textsc{MaxCut} on 7-node graphs. Our circuit depth is also often larger than the previous mentioned research. Since our operation pool consist of single-qubit gates on each qubit and CNOT gates either on neighbouring qubits or between every two qubits in the system, resulting a much larger size of operation pool compared to other research. With these two factors combined, our search space size is generally larger than other QAS research.


However, there are still several hyper-parameters that need to be tuned before the search algorithm can produce satisfying results, which leaves us space for improvement for the automation level of the algorithm. In the future, we would like to investigate the performance of our algorithm under noises, as well as improve the scalability of our algorithm by introducing parallelization to the tree search algorithm when using a quantum simulator. We would also like to introduce more flexible value and/or policy functions into the algorithm.

Overall, our research has shown that MCTS enhanced with combinatorial multi-armed bandit is a very efficient approach to search for quantum circuits for a variety of problems, even when the search space is large. Therefore, it took an important leap towards the widespread applications of variational quantum algorithms on many problems.

\paragraph{Acknowledgement} The authors acknowledge the support from Defense Science Institute. We thank useful discussions and advice from Hanxun (Curtis) Huang. The computational resources were provided by the National Computing Infrastructure (NCI) and Pawsey Supercomputing Center through National Computational Merit Allocation Scheme (NCMAS).



\bibliographystyle{unsrt}
\typeout{} 
\bibliography{reference}

\end{document}